\documentclass[a4paper,11pt]{article}
\pdfoutput=1 

\usepackage{jheppub} 
\usepackage{cancel}
\usepackage[T1]{fontenc} 
\usepackage{comment}
\usepackage{graphicx}
\usepackage{xcolor}
\usepackage{booktabs}
\usepackage{tabularx} 
\usepackage{caption} 

\title{\boldmath A post-inflationary kinetic axion}

\author[a,b]{Enrico Morgante}
\author[a,c]{Riccardo Natale}

\affiliation[a]{Dipartimento di Fisica, Università di Trieste, Strada Costiera 11, I-34151 Trieste, Italy}
\affiliation[b]{INFN, Sezione di Trieste, Via Valerio 2, 34127 Trieste, Italy}
\affiliation[c]{Deutsches Elektronen-Synchrotron DESY, Notkestr. 85, 22607 Hamburg, Germany}

\emailAdd{enrico.morgante@units.it}
\emailAdd{riccardo.natale@desy.de}

\abstract{
We present a novel realization of axion kinetic misalignment, triggered by a Hubble-induced phase transition during a post-inflationary stiff (kination) era. A negative Ricci scalar flips the sign of a non-minimally coupled mass term for a non-minimally coupled complex field $\Phi$, driving its radial mode to large amplitudes via a tachyonic instability. At large $|\Phi|$, higher-dimensional $U(1)$-breaking operators become relevant and impart a kick in the angular direction, generating a conserved $U(1)$ charge that sustains rotation as the symmetry is approximately restored. Because phases randomize across causally disconnected regions, multiple domains with distinct charges form. The subsequent axion potential converts the domain charges into an axion abundance, yielding dark matter even when the net global charge vanishes. We analyze the dynamics through a linear, domain-averaged treatment and identify two thermal histories: (i) Ricci reheating via saxion decays to Higgs bosons; (ii) external reheating with efficient damping of saxion energy by Higgs/fermion scatterings. The mechanism populates regions underabundant in standard misalignment, which are accessible to next generation axion searches.
}

\begin{document} 
\rightline{DESY-25-177}
\maketitle
\flushbottom
\section{Introduction}
The QCD axion and, more in general, axion-like particles (ALPs), constitute a well-motivated dark matter candidate. There are three main ways in which the relic abundance may be generated: via the standard misalignment mechanism~\cite{Preskill:1982cy, Abbott:1982af, Dine:1982ah}, from the decay of topological defects (most notably cosmic strings)~\cite{Gorghetto:2018myk, Gorghetto:2020qws, Gorghetto:2021fsn}, and from scatterings in the thermal plasma, see e.g. Refs.~\cite{Baumholzer:2020hvx, Becker:2025yvb}. In this scenario, the axion field starts coherent oscillations around its potential minimum once its mass exceeds the Hubble rate, yielding a relic abundance set by the initial displacement, the axion mass $m_a$, and the decay constant $f_a$. Over broad regions of parameter space, particularly at low values of $f_a$, the standard mechanism underproduces dark matter.
This region is also the most accessible to experiments which look for the interaction of axions with SM fields, which is controlled by the inverse of the decay constant $f_a^{-1}$. It is therefore important to look for viable DM candidates at low $f_a$, which could be within experimental reach in present or near future experiments.
The kinetic misalignment mechanism, introduced in Refs.~\cite{Co:2019jts, Chang:2019tvx}, offers one possibility: if the axion acquires sufficient early-time kinetic energy, the field rotates in field space, crossing potential barriers and eventually becoming trapped once its kinetic energy redshifts below the barrier height. This trapping can occur well after the standard misalignment condition $H\sim m_a$ is met, so that the resulting axion oscillations are less redshifted, allowing the correct relic density to be achieved in otherwise underabundant regions.

The required large field velocity originates from explicit $U(1)$-breaking at high energies, induced by non-renormalizable operators $V_{\cancel{U(1)}}$. Several implementations of kinetic misalignment have been proposed \cite{Co:2020dya, Kobayashi:2020bxq, Domcke:2022wpb, Co:2023mhe, Lee:2023dtw, Lee:2024bij, Eroncel:2024rpe}. In these scenarios, the saxion attains large field values during inflation, due to quantum fluctuations or to a coupling with the inflaton, and later starts oscillating once the Hubble rate drops below the radial mass. At that stage, the $U(1)$-breaking operator becomes relevant and imparts a kick in the angular direction, generating an axion field velocity; if large enough, this delays the onset of oscillations and realizes a kinetic misalignment scenario.
The same model has found a number of different applications in the literature, including baryogenesis, gravitational waves and modified cosmic history with a kination phase.~\cite{Co:2019wyp, Co:2020jtv, Madge:2021abk, Co:2021rhi, Harigaya:2021txz, Co:2021lkc, Co:2021qgl, Gouttenoire:2021wzu, Gouttenoire:2021jhk, Barnes:2022ren, Co:2022qpr, Co:2022aav, Barnes:2024jap, Bodas:2025wef, Co:2025lrd, Eroncel:2025qlk, Bodas:2025eca}.

In this work, we propose a new realization of kinetic misalignment, in which the large field displacement is due to a negative Hubble-induced mass term at the end of inflation. 
In our model, the saxion couples non-minimally to the Ricci scalar $R=3(1-3w)H^2$, which turns negative during a stiff era with $w>1/3$ at the end of inflation.
We can assume, for simplicity, that the equation of state is $w\approx1$, leading to a kination era. This occurs for example, if at the end of slow-roll the inflaton encounters a steep potential and its kinetic energy dominates~\cite{Spokoiny:1993kt, Peebles:1998qn, Akrami:2017cir, Dimopoulos:2019ogl}.
At the inflation–kination transition, the Ricci scalar changes sign, triggering spontaneous symmetry breaking and generating a new minimum at large field values. The resulting tachyonic instability drives the radial mode far from the origin, making the $U(1)$-breaking operator relevant and imparting a kick in the angular direction. As the radial mode subsequently redshifts back, the $U(1)$ symmetry is approximately restored, and the later dynamics are determined by $U(1)$ charge conservation.
Similar scenarios have been considered before in the literature under the name of Hubble-induced phase transitions~\cite{Bettoni:2018utf, Bettoni:2019dcw, Bettoni:2021zhq, Bettoni:2024ixe}.
A key aspect is that of reheating, for which we describe two possibilities: (i) the radial energy density is transferred to the Standard Model plasma, effectively reheating the Universe via the Ricci reheating mechanism~\cite{Opferkuch:2019zbd, Laverda:2023uqv, Figueroa:2024asq}; (ii) reheating proceeds by another mechanism, while the radial energy is dissipated through scattering with the plasma.

Our setup differs from conventional kinetic misalignment scenarios, as the radial mode reaches large field values after inflation due to a tachyonic instability. This inevitably induces the growth of perturbations and, after the kick, an inhomogeneous axion field velocity across the Universe. The situation closely parallels post-inflationary misalignment, where topological defects such as cosmic strings and domain walls form. Here we perform a linear analysis of the dynamics, assuming that after the kick several domains with different axion velocities emerge, and that gradient energy is negligible within each domain. We discuss the validity of this assumption below, but we plan to test them with lattice simulations in future work.

Our paper is organized as follows.
In Sec.~\ref{sec:Initial Field Dynamics} we construct the model and describe the dynamics leading to a successful kinetic misalignment. In Sec.~\ref{sec:inhomogeneous universe} we discuss the small scale behaviour after inflation, comparing our scenario to the more common post-inflationary axion misalignment. Section~\ref{sec:rehating} deals with how reheating can be achieved in our model. In Sec.~\ref{sec:parameter space} and~\ref{sec:Axion Dark Matter Abundance} we show the allowed parameter space and the calculation of the DM abundance, respectively. A second scenario, in which reheating originates from a different sector, is discussed in Sec.~\ref{sec:spectator saxion}. Section~\ref{sec:conclusions} contains our conclusions.

\section{Initial Field Dynamics}\label{sec:Initial Field Dynamics}
\subsection{The model}

We introduce the model, writing down its action in the Jordan frame
\begin{equation}
    S=\int \text{d}^4x\sqrt{-g}\left[\frac{M_{\text{Pl}}^2}{2}R+\mathcal{L}_{\text{inf}}-g^{\mu\nu}\partial_\mu\Phi\partial_\nu\Phi^{\dagger}-\xi R|\Phi|^2-V_{U(1)}(|\Phi|^2)-V_{\cancel{U(1)}}(\Phi,\Phi^{\dagger})\right],
    \label{action model}
\end{equation}
where $R=3(1-3w)H^2$ is the Ricci scalar, $\mathcal{L}_{\text{inf}}$ is the inflaton sector Lagrangian, $\Phi$ is a complex scalar field charged under a global $U(1)$ symmetry, non-minimally and non-conformally coupled to gravity. The metric $g_{\mu\nu}$ is assumed to be the flat Friedmann-Lema\^itre-Robertson-Walker one
\begin{equation}
    \text{d}s^2=-\text{d}t^2+a(t)^2\delta_{ij}\mathrm{d}x^i\mathrm{d}x^j.
\end{equation}
During inflation, because $R\approx12H_I^2$, $\Phi$ acquires a large mass term $m^2\approx12\xi H_I^2$, preventing the spontaneous breaking of $U(1)$ symmetry and the formation of large quantum fluctuations.

We assume that inflation ends with a \textit{stiff} era with equation of state $w>1/3$, lasting for a few e-folds. In particular, in presenting our results we will be mostly concerned with the \textit{kination} eos $w=1$, although our results will be valid more in general. Formulas for generic $w$ will be presented in appendix.
During this era, the Ricci scalar turns negative, providing this time a tachoynic mass to $\Phi$,  $m^2\propto -H^2$, driving the field to large values. At this point the $U(1)$ symmetry is spontaneously broken and the axion is established as a Goldstone boson. In order to generate a net $U(1)$ charge, we introduce an explicit $U(1)$ breaking represented by the higher dimensional operator $V_{\cancel{U(1)}}$ in Eq.~(\ref{PQ breaking potential}) below. If $\Phi$ is large enough, this term becomes relevant and impart a kick in the angular direction, resulting in a non-zero axion field velocity.

In this section we study in detail the dynamics of $\Phi$, specifically of its radial and angular mode, by numerically solving the homogeneous equations of motion of these two degrees of freedom.

We start by writing down the e.o.m of $\Phi$
\begin{equation}
    \ddot\Phi - \frac{\nabla^2\Phi}{a^2} +3H\dot\Phi+\xi R\Phi+\frac{\partial V}{\partial\Phi^\dagger}=0,
    \label{equation Phi}
\end{equation}
where we choose
\begin{equation}
        V_{U(1)}=\lambda\left(|\Phi|^2-\frac{f_a^2}{2}\right)^2,\quad
        V_{\cancel{U(1)}}=2^{\frac{n}{2}}\frac{A\Phi^n}{n M_{\mathrm{Pl}}^{n-3}}+\mathrm{h.c.}.
    \label{PQ breaking potential}
\end{equation}
We apply the decomposition 
\begin{equation}\label{polar decomposition}
    \Phi=\frac{S}{\sqrt{2}}e^{i\theta}.
\end{equation}
where we recognize the angular degree of freedom as the axion and $S$ is the radial mode, which we will refer to as the saxion. The $U(1)$ conserving potential in Eq.~(\ref{PQ breaking potential}) has a minimum in $S=f_a$, with a saxion mass in the vacuum $m_S^2=2\lambda f_a^2$.
Equation~\eqref{equation Phi} is thus split into two real scalar equations for $S$ and $\theta$
 \begin{subequations}
    \begin{align}
        \ddot S-\frac{\nabla^2S}{a^2}+3H\dot S +(\xi R-\lambda f_a^2)S+\lambda S^3+2A\frac{S^{n-1}}{M_{\mathrm{Pl}}^{n-3}}\cos(n\theta)&= S\dot\theta^2 -S\frac{(\vec\nabla\theta)^2}{a^2} \label{saxion}\\
        \ddot\theta-\frac{\nabla^2\theta}{a^2}+3H\dot\theta-2 A\frac{S^{n-2}}{M_{\text{Pl}}^{n-3}}\sin(n\theta)&=-2\frac{\dot S}{S}\dot\theta+\frac{2}{a^2}\frac{\vec\nabla S}{S}\cdot\vec\nabla\theta \,.\label{axion}
    \end{align}
\end{subequations} 
Here we included the gradient terms for completeness, although we are going to neglect them in the following. This approximation is valid on scales smaller than the tipical correlation lenght of the fields $S,\theta$, and as long as resonance effects can be neglected. We will further examine this issue below.

\subsection{Saxion dynamics at the end of inflation}
Inflation ends, in our scenario, with a stiff era, in which $R = 3(1-3w)H^2<0$. We assume that $\lambda f_a^2,\, \lambda S_I^2\ll-\xi R$, and Eq.~\eqref{saxion} can be  approximated to
\begin{equation}
    \ddot S+3H\dot S+\xi RS\simeq0.
    \label{growing eq}
\end{equation}
In this section we will take for simplicity $w=1$ after inflation, resulting in a perfect kination era. In App.~\ref{sec:appendix tachyonic growth} we display the full dependence on $w$. The last term in Eq.~\eqref{growing eq} turns negative, since $R=-6H^2$, thus the solution has a tachyonic growth described by
\begin{equation}
    S(t)=bS_I\left(\frac{H_I}{H}\right)^\alpha
    \label{growing solution}
\end{equation}
with $\alpha=\sqrt{2\xi/3}$ and $b\approx1$ for $w=1$.
We choose as initial conditions the RMS of fluctuations computed at the end of inflation, as derived in App.~\ref{NMC}:
\begin{align}
    S_I&=\frac{H_I}{\sqrt{6}\pi (12\xi)^{1/4}},\\
    \dot S_I&=\sqrt{6\xi}H_IS_I.
\end{align}
The tachyonic growth lasts until the saxion field is stopped by self-interaction through the quartic potential. This happens when the growing solution of Eq.~(\ref{growing solution}) reaches the minimum of the potential, controlled by $V_\mathrm{eff}\supset \xi R S^2/2 + \lambda/4 S^4$. Equivalently, the stopping condition can be obtained by assuming that the approximation made for Eq.~\eqref{growing eq} breaks down, resulting in
\begin{equation}
    -\xi R_\text{max} \approx \lambda S_{\text{max}}^2\,,
    \label{stopping condition}
\end{equation}
where we will use from now on the subscript ``max'' for quantities computed when the saxion reaches its maximum value. We explicitly compute the Hubble scale $H_{\text{max}}$ and the saxion value $S_{\text{max}}$ combining Eq.~\eqref{stopping condition} with Eq.~\eqref{growing solution}, getting 
\begin{subequations}
    \begin{align}
        H_{\text{max}}&\simeq H_I\left(\frac{\lambda}{72\sqrt{3}\pi^2\xi^{3/2}}\right)^{\frac{1}{2+2\alpha}},\label{Hmax}\\
        S_{\text{max}}&\simeq\sqrt{\frac{6\xi}{\lambda}} H_{\text{max}} \simeq S_I\left(\frac{72\sqrt{3}\pi^2\xi^{3/2}}{\lambda}\right)^{\frac{\alpha}{2+2\alpha}}\,.\label{Smax}
    \end{align}
\end{subequations}

\begin{figure}
    \centering
    \includegraphics[scale=0.5]{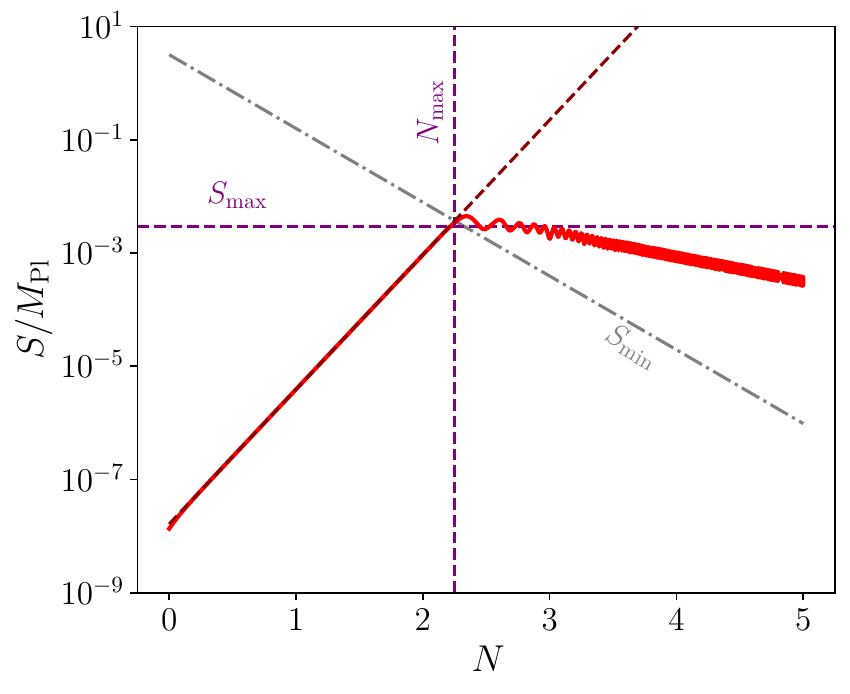}
    \includegraphics[scale=0.5]{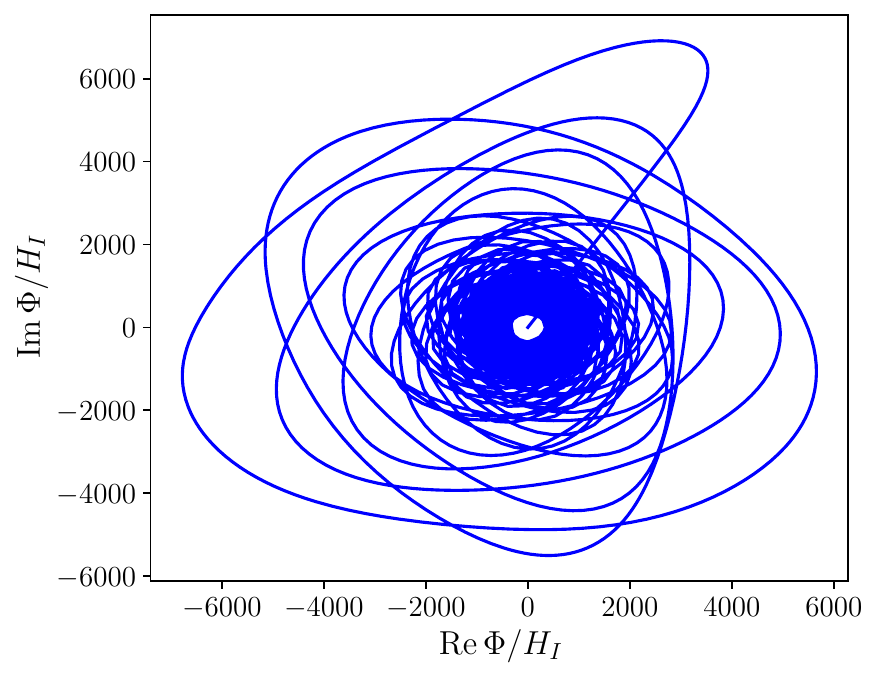}
\caption{\textit{Left:} Evolution of the saxion field $S$. The numerical result is shown in red, the analytic solution of Eq.~\eqref{growing solution} is displayed with a dashed brown line, showing a good agreement during the tachyonic growth. The two dashed purple lines represent Eqs.~\eqref{Hmax} and~\eqref{Smax}, while the position of the minimum Eq.~\eqref{Smin} is displayed in grey.\\
On the right, the same evolution in the complex plane, where the rotating dynamics of $\Phi$ is showed. The parameters used are $H_I=10^{12}\,\text{GeV},\,f_a=10^{10}\,\text{GeV},\,\lambda=10^{-12},\,\xi=5,\,n=7$.}
    \label{Saxion sol}
\end{figure}

In Fig.~\ref{Saxion sol}, we show the evolution of the saxion field $S$ during the phase of tachionic growth and the onset of the oscillating phase, for one example choice of parameters. The numerical evolution is shown in red, while the dashed purple lines show the value of $S_\mathrm{max}$ and the number of efolds $N_\mathrm{max}$ at which this is obtained, showing a good agreement with the numerical results.

When the tachionic growth is halted, the $U(1)$ breaking term becomes relevant and, if $\theta$ is not perfectly aligned to zero, imparts a kick in the angular direction, starting a rotational motion within the quartic potential. 

\subsection{Axion dynamics and generation of a $\text{U}(1)$ charge}
To study the dynamics of the kick we focus now on the axion field. It's useful to introduce the $U(1)$ Noether charge
\begin{equation}
    n_{\theta}=i(\Phi\dot\Phi^{\dagger}-\Phi^{\dagger}\dot\Phi)=S^2\dot\theta \,,
\end{equation}
and we recast Eq.~\eqref{axion} in the following form
\begin{equation}
    \dot n_{\theta}+3Hn_{\theta}=-\frac{\partial V_{\cancel{U(1)}}}{\partial \theta}\,.
    \label{PQ charge}
\end{equation}
Without a $U(1)$ breaking term it reduces to
\begin{equation}
    \frac{\text{d}}{\text{d}t}(a^3n_{\theta})=0
    \label{PQ charge conserv}
\end{equation}
and the charge is conserved in a comoving volume. However, with the presence of $V_{\cancel{U(1)}}$, if the axion is not perfectly aligned to zero, when the radial mode reaches high values, our field will experience a gradient in the angular direction, so the axion will acquire a non zero velocity. The quantity $n_{\theta}$ produced after the kick can be estimated by integrating Eq.~\eqref{PQ charge},  as we do explicitly in App.~\ref{Axion kick}, resulting in the following axion velocity
\begin{equation}
   \dot\theta_{\text{max}}=\beta \sin(n\theta)\frac{AS_{\text{max}}^{n-2}}{M_{\text{Pl}}^{n-3}H_{\text{max}}} 
    \label{axion velocity theta}
\end{equation}
and a $U(1)$ charge
\begin{equation}
   a^3 n_\theta = a^3_\mathrm{max} \beta \sin(n\theta)\frac{AS_{\text{max}}^n}{M_{\text{Pl}}^{n-3}H_{\text{max}}}
    \label{axion velocity charge}
\end{equation}
where $\beta$ is a $\mathcal{O}(1)$ number, the precise value of which is given in App.~\ref{Axion kick}.

After the kick, due to Hubble friction, the saxion field redshifts back to $f_a$, following a radiation-like equation of state
\begin{equation}
    \langle w_S\rangle = 1/3,\qquad \langle\rho_S\rangle\propto a^{-4}, \qquad \langle S\rangle \propto a^{-1}.
    \label{saxion in the quartic}
\end{equation}
where the average is on a complete rotation.%
\footnote{It should be stressed here that the orbit of the field in the quadratic plus quartic potential is not, in general, elliptic, nor does it follow closed orbits. The definition of the average velocity is thus somewhat imprecise, although the implicit arbitrariness in this definition is smaller than the uncertainty implicit in our order of magnitude estimates. We refer to App.~\ref{asymmetry factor} for further details.}
This can be understood, if we look at the saxion-axion system in Cartesian basis $\Phi=X+iY$, so that the rotating complex field can be decomposed into two real fields $X$ and $Y$ oscillating in a quartic potential. From the virial theorem, it results that $\rho_{\text{kin}}=2\rho_{\text{pot}}$, which leads to $w_S=1/3$.

The redshift $S\propto a^{-1}$ makes the term $V_{\cancel{U(1)}}\propto a^{-n}$ irrelevant soon after the kick. The axion dynamics is then fixed by charge conservation as in Eq.~\eqref{PQ charge conserv} and we have $a^3n_{\theta}=\text{const.}$ From Eq.~\eqref{saxion in the quartic} and the definition of $n_{\theta}$ we obtain
\begin{equation}
    \langle\dot\theta\rangle=\dot\theta_{\text{max}}\left(\frac{a_{\text{max}}}{a}\right).
\end{equation}
where the average is taken on a full orbit. Neglecting Hubble friction, we can assume a quasi elliptical motion in the $S,\theta$ plane. The eccentricity of the orbit can be measured by the asymmetry factor $\epsilon$, which we define as~\cite{Co:2020jtv}
\begin{equation}\label{epsilon def}
    \epsilon \equiv \frac{n_\theta}{n_S} = \frac{n_{\theta}}{S^2\sqrt{V'/S}} \leq 1\,,
\end{equation}
where the quantities on the right hand side are computed at the point of furthest distance from the center along the field's orbit, and we defined $n_S \equiv S^2\sqrt{V'/S}$.
A perfectly circular orbit is obtained when the centrifugal and centripetal forces balance, that happens for $\dot\theta = \sqrt{V'/S}$, ie $\epsilon=1$.
An efficient kick mechanism requires $\epsilon\approx 1$.
As we will explain in Section \ref{sec:inhomogeneous universe}, in our model all possible realizations of $\epsilon$ are present, in separate regions within the universe. Therefore, requiring that $\epsilon \approx 1$ should be understood as a spatial average within our universe.
Smaller $\epsilon$ correspond to a more elongated orbit, with non-harmonic oscillations in the radial direction that could lead to a parametric resonance effect~\cite{Co:2020dya, Fedderke:2025sic}. If this case, perturbations in the saxion and axion field grow depending on the shape of the orbit, with an efficiency that grows for small $\epsilon$. Although this effect should not affect dramatically our prediction of the DM abundance, it will have some interesting consequence due to the production of large momentum axions. We plan to study this effect in a future work, in which we will address inhomogeneities in more detail.

In the calculation of $\epsilon$, one can neglect the term coming from the non-minimal coupling in $V$ (see Eq.~\eqref{NMC energy} for the full expression), which decreases in time during the kination phase, and retain only the quadratic and quartic terms.
In App.~\ref{asymmetry factor} we further examine the relation between $\epsilon$ and the eccentricity $e$ of the orbit.

\subsection{Axion/saxion evolution and early cosmology}
In this section, we briefly describe the evolution of the axion and saxion fields after the $U(1)$ charge is generated. We envisage two main cosmological scenarios, depending on which field is responsible for reheating after inflation.

\subsubsection{Saxion reheating the universe}
After reaching the maximum, the quartic term in the saxion potential quickly dominates over the other components of $T^{00}$. The energy density of the saxion scales as $a^{-4}$ until it relaxes to the minimum of the potential $S= f_a$. The universe is initially dominated by the inflaton sector, which by assumption undergoes a stiff era with $a^{-3(1+w)}\sim a^{-6}$. The kination era ends when the saxion energy density dominates over the inflaton one.
The Hubble parameter at the time when kination ends can be estimated by taking
\begin{equation}
    3 M_{\text{Pl}}^2 \,H_\mathrm{ke}^2 = \frac{\lambda}{4} S_\mathrm{ke}^4\,.
\end{equation}
Defining $e^{\Delta N_\mathrm{ke}} = a(t_{\mathrm{ke}})/a_\mathrm{max}$, we obtain
\begin{equation}
    e^{\Delta N_\mathrm{ke}} = \frac{\lambda^{1/2}}{\sqrt{3}\,\xi} \frac{M_\mathrm{Pl}}{H_\mathrm{max}}
\end{equation}
and
\begin{equation}
    H_\mathrm{ke} = \frac{3\sqrt{3}\,\xi^3}{\lambda^{3/2}}\frac{H_\mathrm{max}^4}{M_\mathrm{Pl}^3} \,.
    \label{Hke}
\end{equation}

\begin{figure}[t]
    \centering
    \includegraphics[width=1\linewidth]{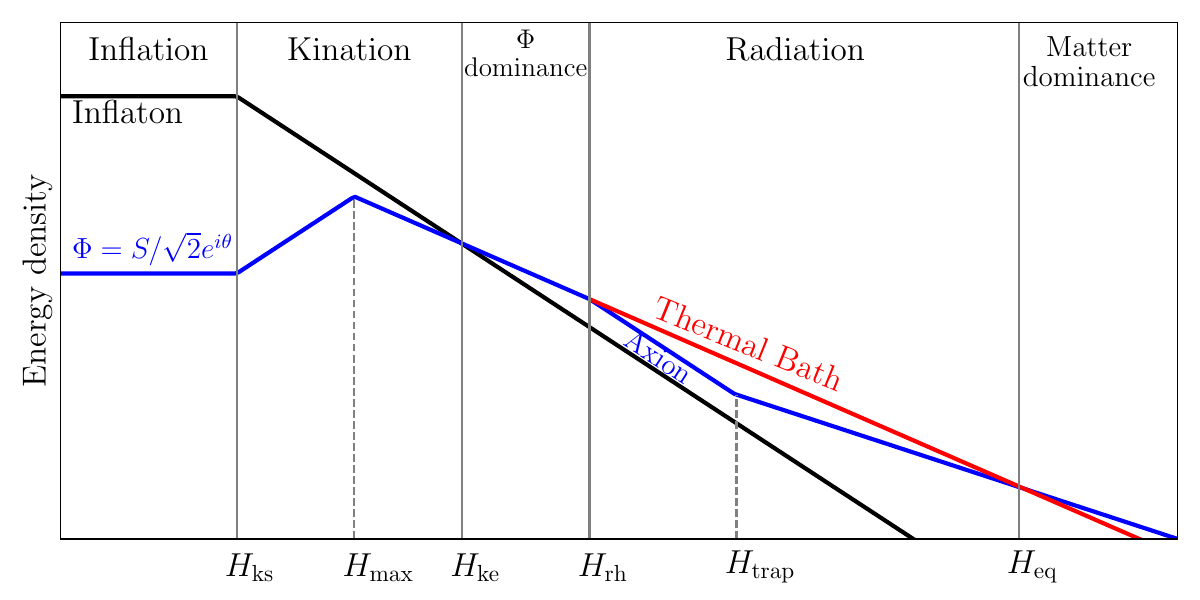}
    \caption{
    Evolution of the components of the energy density, in the scenario in which the saxion reheats the universe. At the end of inflation, a period of kination starts at $H_{\text{ks}}$, during which the field $S$ grows until it reaches a maximum value at $H_\mathrm{max}$, and then scales as $\rho_S\sim a^{-4}$. At $H_\mathrm{ke}$ it overtakes the inflaton energy density. At $H_\mathrm{rh}$ the saxion decays and reheats the thermal bath. The energy in the axion condensate scales as $a^{-6}$ (kination) until the field gets trapped in the potential at $H_\text{trap}$, and the dark-matter behaviour begins.
    }
    \label{fig:energy evolution}
\end{figure}

At this point, the energy of the saxion field must be transferred to the SM plasma, in a saxion-driven reheating, the details of which will be discussed below.
A sketch of the evolution of the energy density is shown in Fig.~\ref{fig:energy evolution}.
We discuss how this can be implemented in Sec.~\ref{sec:rehating}. The relevant parameter space for this scenario is illustrated in Sec.~\ref{sec:parameter space}, while the relic axion abundance is computed in Sec.~\ref{sec:Axion Dark Matter Abundance}.

\subsubsection{Saxion as a spectator field}

\begin{figure}
    \centering
    \includegraphics[width=1\linewidth]{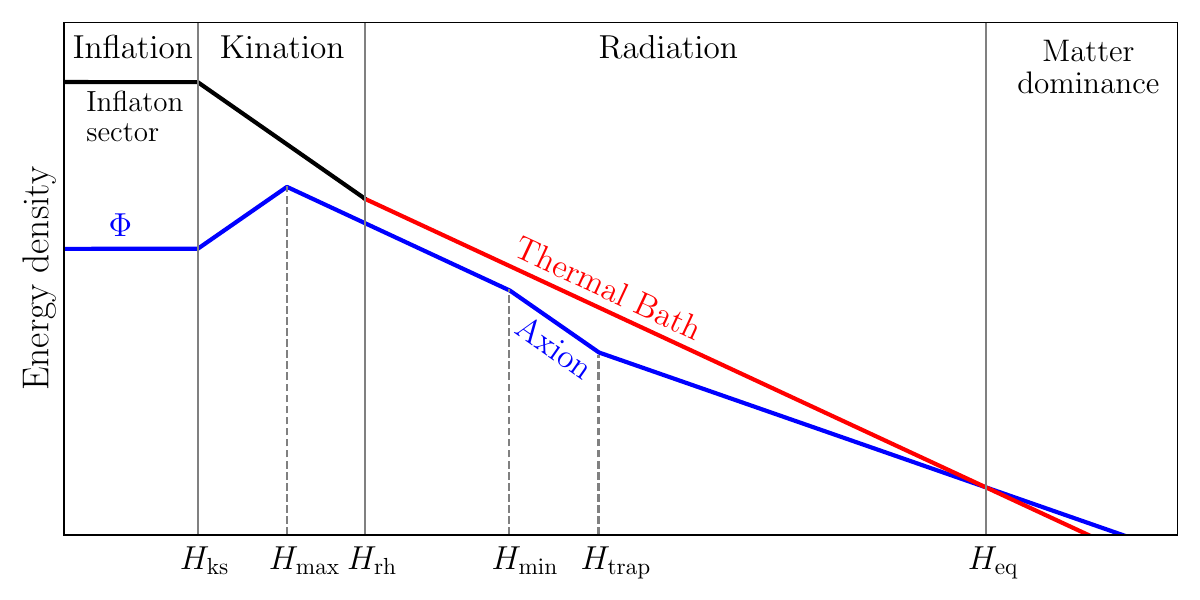}
    \caption{Same plot as the one showed in Fig.~\ref{fig:energy evolution}, but in this scenario the $\Phi$ does not dominate the energy density and reheating is achieved via some other mechanism.}
    \label{fig:scenario_spectator}
\end{figure}

As an alternative to the scenario discussed above, we consider the case in which the field $\Phi$ is a spectator which does not dominate the energy density of the Universe until matter-radiation equality. Reheating is controlled by the inflaton or by some other field which does not couple to $\Phi$. For simplicity, we assume that reheating takes place after the saxion has reached its maximum value $S_\mathrm{max}$, but before it comes to dominate the energy density. The evolution of the energy density is displayed in Fig.~\ref{fig:scenario_spectator}.
This possibility is analysed in more detail in Sec.~\ref{sec:spectator saxion}.


\section{Axion inhomogeneities}\label{sec:inhomogeneous universe}

In the results presented in Sec.~\ref{sec:Initial Field Dynamics}, we neglected gradient terms and assume a perfectly homogeneous field. However, we expect substantial inhomogeneities to arise in our scenario, from at least three sources. Firstly, the kinetic misalignment mechanism features non-trivial perturbation dynamics, including axion fragmentation close to trapping, which can alter the early evolution of the field and the final dark matter abundance \cite{Fonseca:2019ypl, Morgante:2021bks, Eroncel:2022vjg, Eroncel:2022efc, Eroncel:2025qlk, Bodas:2025eca, Fasiello:2025ptb}.
Secondly, the fast Hubble-induced phase transition will generate a tachyonic instability which affects any initially small perturbation.
Finally, and most importantly, kinetic misalignment proceeds independently in separate patches of the Universe, similarly to the more standard post-inflationary misalignment scenario.

In this section we will comment on these small-scale inhomogeneities and the regime of validity of our computation. The full evolution is clearly non-linear, and a dedicated lattice study is required that we postpone it to future works.

\subsection{Domains}

When inflation ends and the equation of state of the universe transitions to kination, and the $U(1)$ symmetry breaks spontaneously thanks to the Hubble-induced negative mass term. At this moment, different regions of the universe select different $\theta$ angles within the interval $(0,2\pi)$. From Eq.~\eqref{axion velocity charge} we can see that the initial angle $\theta$ determines the strength and direction of the kick in the angular direction. Thus, having a universe with a random distribution of $\theta$ results in a randomization of $\dot\theta$ after the kick is imparted. At this point, we expect the resulting universe to be composed of several regions in which the value of $n_{\theta}$ differs in value and sign. In other words, in some of these domains the axion field is rotating clockwise, and in others counterclockwise, with some boundaries where the rotation is halted for continuity of the field. The size of each domain should be set by Hubble at the kick
\begin{equation}
    \ell_{\text{dom}}\sim\alpha_{\text{dom}} H_{\text{max}}^{-1}\,.
\end{equation}
Although the evolution is clearly non-linear, we consider the dynamics described in Sec.~\ref{sec:Initial Field Dynamics} to be valid locally, specifically inside each domain. For this reason, we define 
\begin{equation}
    n_{\theta,\text{dom}}=S^2\dot\theta_{\text{dom}}\propto\sin{(n\theta_{\text{dom}})}
\end{equation}
to be the $U(1)$ charge inside a specific domain. Since $S_{\text{max}}$ does not depend on the angle, we take it to be constant in each domain,%
\footnote{Note that perturbations of $S$ are generated during the tachyonic phase and contribute to inhomogeneities of $n_{\theta,\text{dom}}$. However, they will be suppressed by $M_\mathrm{Pl}$, and can be neglected.}
unlike $\dot\theta$ that inherits its dependence on $\theta_{\text{dom}}$ as showed in \eqref{axion velocity theta}. Inside each domain the gradient energy density is subdominant compared to kinetic energy density, we can estimate the ratio between gradient and kinetic energy as
\begin{equation}
    \frac{S^2(\nabla\theta)^2}{S^2\dot\theta^2}\sim\left(\frac{k_{\text{dom}}}{\dot\theta_{\text{max}}}\right)^2\sim\left(\frac{H_{\text{max}}}{\sqrt{\lambda}S_{\text{max}}}\right)^2\sim\frac{1}{6\xi}\,.
\end{equation}
This ratio is smaller than one if $\xi\gg1$. Under this assumption, perturbations are negligible within the domain and $a^3n_{\theta,\text{dom}}$ is conserved.
On the other hand, between two domains with opposite rotation we have
\begin{equation}
    \Delta\theta\sim2\dot\theta_{\text{dom}}t,\label{excursion boundary}
\end{equation}
resulting in a large $|\partial_i\theta|$ along the boundaries.

The inhomogeneities in the DM density are not, in principle, at odds with cosmological observations, because they appear on a very small scale only. In this sense, our model is similar to the usual post-inflationary axion misalignment: in each point in space, the axion selects a different initial value at the time of symmetry breaking, and the DM abundance is computed by averaging over the $2\pi$ range.
On top of this, adiabatic perturbations are transferred to the axion by the single clock represented by the inflaton and later by the temperature of the plasma. Perturbations of the inflaton field control the onset of kination and thus the spontaneous symmetry breaking from which the axion emerges, again similarly to the standard post-inflationary scenario.

A similar set up has been proposed in Ref.~\cite{Bettoni:2018utf}, where the Hubble-induced phase transition is used to impart a rotation to an Affleck-Dine field in order to reproduce the observed baryon asymmetry. As the authors comment, a global non zero $B-L$ asymmetry can be generated only for non $Z_n$-symmetric operators with odd $n$ and complex coupling, such that in the universe there will be a different amount of clockwise and anti-clockwise domains.
In our story, the charge $n_\theta$ generated at the kick will be converted into axion number density $n_a$ regardless the sign of the initial charge. For this reason, even a zero global charge, with a non-zero variance, will efficiently produce a non-zero axion abundance once the axion obtains a mass.
Although in principle the small scale dynamics could partly wash out the charge locally, energy conservation arguments imply that $n_\theta$ cannot be erased. We postpone to a future work a detailed study of this effect.

\subsection{Topological defects}
Because of the phase transition occurring after inflation, the Kibble mechanism~\cite{Kibble:1976sj, Vilenkin:1981kz, Sikivie:1982qv, Vilenkin:1982ks, Sikivie:2006ni} is active in our model, and due to the spontaneous breaking of a $U(1)$ symmetry, long strings are expected to form. We assume that their evolution is close to the standard post-inflationary scenario and their impact on the dark matter abundance should be negligible for small $f_a$. We will come back to this point in Section \ref{sec:parameter space}.

Moreover, another source of topological defects is expected. Along the boundaries between two domain with opposite in sign $U(1)$ charges, the axion field reaches large excursions following Eq.~\eqref{excursion boundary} and when $\Delta\theta\approx\mathcal{O}(2\pi)$ string loops will form. The evolution of such loops is very different compared to the usual long strings produced by the Kibble mechanism. Ref.~\cite{Fedderke:2025sic} studies the formation of domains due to parametric resonance in the case $\epsilon\ll1$. Their results show the formation of loops along the boundaries, where they are also confined, leading to a rapid decay without ever reaching a scaling regime. We postpone to a future work the study of these loops and the possible emission of gravitational waves.

\section{Reheating from saxion decay}\label{sec:rehating}

In this section we discuss one scenario for reheating, in which the saxion is unstable and can decay into the SM and possibly other BSM fields. If at the time of decay, no thermal bath is present, the saxion will be the actual reheaton field of our universe starting the radiation era, in analogy to the scenario of Ricci Reheating~\cite{Opferkuch:2019zbd, Bettoni:2021zhq, Laverda:2023uqv, Figueroa:2024asq, Bettoni:2024ixe} . In order to describe this decay process, we introduce a decay rate $\Gamma$ and solve the equations of motion of saxion and axion and the resulting thermal bath. 

\subsection{Decay via Yukawa interactions}
A possible interaction, typical of KSVZ axion model, is with fermions through a Yukawa term 
\begin{equation}
    \mathcal{L}_{\text{}}\supset yS\bar\chi\chi,
    \label{yukawa coupling}
\end{equation}
These fermions can be charged under SU(3) color, so after their production via saxion decay they will produce the SM plasma via strong interactions. The perturbative decay $S\rightarrow\bar\chi\chi$ is given by
\begin{equation}
    \Gamma_{S\bar\chi\chi}=\frac{y^2}{8\pi}m_S \left(1-\frac{4m_\chi^2}{m_S^2}\right)^{\frac{3}{2}}.
\end{equation}
In order for this decay channel to be open we need that $m_S>2m_{\chi}$. However because the large saxion vev, fermions get a mass from the term \eqref{yukawa coupling}
\begin{equation}
    m_{\chi,\text{eff}}=m_{\chi}+yS,
\end{equation}
so asking for the channel to be open it requires $4y^2S^2<3\lambda S^2$
\begin{equation}
  \frac{y}{\sqrt{\lambda}}<\sqrt{\frac{3}{4}}.
  \label{upper bound y}
\end{equation}
Moreover, the presence of another decay rate has to be taken into account. Indeed, the saxion is coupled to the axion via the kinetic term, giving a decay channel for the saxion provided by
\begin{equation}
    \mathcal{L}_{\text{int}}=\frac{s}{\langle S\rangle}(\partial_{\mu}a)^2,
\end{equation}
where we expanded $S=s+\langle S\rangle$, with $\langle S\rangle$ the saxion vev of the condensate and $s$ are perturbations around it. From this interaction the decay rate is
\begin{equation}
    \Gamma_{\text{axion}}\simeq\frac{m_S^3}{32\pi S^2}\simeq\frac{3^{3/2}\lambda^{3/2}S}{32\pi}.
    \label{saxion-into-axion decay}
\end{equation}
If this decay is the preferred one, the saxion energy density will be transferred into a hot population of stable axions and the Hot Big Bang scenario will never be realized. To avoid this, reheating has to happen before the decay in Eq.~\ref{saxion-into-axion decay} becomes relevant.
To reheat via fermions, we impose $\Gamma_{\text{axion}}\ll\Gamma_{S\bar\chi\chi}$, which means that
\begin{equation}\label{eq:Yukawa vs axions decay}
    \sqrt{\frac{3}{4}}\ll\frac{y}{\sqrt{\lambda}}.
\end{equation}
However, this bound is incompatible with the requiring that the decay channel into fermions is open, Eq.~\eqref{upper bound y}, for this reason we have to exclude this scenario.

The coupling to fermions allows, at one loop, the decay of the saxion into gluons and photons, controlled by
\begin{equation}
    \Gamma_{S\gamma\gamma} \sim y^2 \alpha^2\frac{m_S^3}{m_{\chi,\mathrm{eff}}^2}
    \approx \alpha^2\frac{m_S^3}{S^2}
\end{equation}
times numerical factors including fermion charges and multiplicities. A similar formula holds for gluons. Since the parametrics of this decay rate is the same as that into axions, the latter can not be suppressed with respect to it, thus again we cannot rely on this decay channel to deplete the saxion abundance.

\subsection{Decay via Higgs portal}
Another possible realization is via a coupling to the SM Higgs of the form
\begin{equation}\label{Higgs coupling}
    \mathcal{L}\supset\lambda_{\text{SH}}\left(S^2-f_a^2\right)H^{\dagger}H \,,
\end{equation}
where we tune the interaction term with $f_a^
2$ in order to avoid a large Higgs mass from the late time saxion vev.%
\footnote{Note that this tuning can be actually included in the usual tuning of the hierarchy problem.}
This term could be avoided were $\lambda_\mathrm{SH}$ small enough, but it turns out that this is not viable in our parameter space.
This coupling induce a decay channel $S\rightarrow H^{\dagger}H$ with a decay rate
\begin{equation}
    \Gamma_{SH^{\dagger}H}=\frac{\lambda_{\text{SH}}^2\langle S\rangle^2}{2\pi m_S}\sqrt{1-\frac{4m_H^2}{m_S^2}},
\end{equation}
where for large saxion vevs $\langle S\rangle\gg f_a$, $m_S=\sqrt{3\lambda}\langle S\rangle$ and $m_H\simeq\sqrt{\lambda_{SH}}\langle S\rangle$, which leads to
\begin{equation}
    \Gamma_{SH^{\dagger}H}=\frac{\lambda_{\text{SH}}^2\langle S\rangle}{2\pi \sqrt{3\lambda}}\sqrt{1-\frac{4\lambda_{SH}}{3\lambda}},
\end{equation}
The channel is open for $3\lambda>4\lambda_\text{SH}$. However, also here we have to ask for this channel to be open before saxion-into-axion decay. We see that the decay into Higgs is the dominant one if $3\lambda<2\lambda_\text{SH}$ and this is again in contrast by the condition imposed above. However, the process can still happen when the saxion is around the minimum and, because of the tuning on the Higgs mass, the saxion may be much heavier than the Higgs and the decay can happen. For $S=f_a$ the decay rate becomes 
\begin{equation}
    \Gamma_{Shh}=\frac{\lambda_{\text{SH}}^4f_a}{2\pi\sqrt{2\lambda}}\sqrt{1-\frac{2m_h^2}{f_a^2\lambda}} \,,
    \label{Higgs decay rate}
\end{equation}
where $m_h=125.35\,\text{GeV}$ and assuming $\lambda>2m_h^2/f_a^2$ this channel is open. 

In order to check the efficiency of the reheating we numerically solve the equations of motion for the energy density. After the saxion reaches the minimum the equations are
\begin{align}
    \dot{\rho}_S&=-(3H+\Gamma)\rho_S,\\
    \dot{\rho}_{\theta}&=-6H\rho_{\theta},\\
    \dot{\rho}_{\text{rad}}&=-4H\rho_{\text{rad}}+\Gamma\rho_{\text{rad}}.
\end{align}
Solving this set of equations we see that reheating can be approximated as instantaneous at the moment when the saxion starts oscillating around the minimum $f_a$. The time when $S\simeq f_a$ can be computed using the scaling of Eq.~\eqref{saxion in the quartic}. Defining $a_\mathrm{min}$ the scale factor at this time we can write
\begin{equation}\label{saxion recheas f}
f_a=S_{\text{max}}\left(\frac{a_{\text{max}}}{a_\mathrm{min}}\right).
\end{equation}
In terms of Hubble scale the saxion reaches the minimum $S=f_a$ when
\begin{equation}
    H_\mathrm{min}=H_{\text{ke}}\left(\frac{f_a}{S_{\text{max}}}\right)^2\left(\frac{H_{\text{max}}}{H_{\text{ke}}}\right)^{\frac{4}{3(1+w)}},
\end{equation}
which, for any $w$ in the stiff region $1/3<w<1$, simplifies to
\begin{equation}
    H_\mathrm{min}=\frac{\sqrt{\lambda}f_a^2}{2\sqrt{3} M_{\text{Pl}}}\,.
\end{equation}
Assuming then an instantaneous reheating we have
\begin{equation}
    T_{\text{rh}}\approx 10^6\,\text{GeV}\,\left(\frac{100}{g}\right)^{1/4}\left(\frac{\lambda}{10^{-10}}\right)^{1/4}\left(\frac{f_a}{10^9\,\text{GeV}}\right).
    \label{Reheating Temperature}
\end{equation}

Note that, even though the Higgs decay channel is closed before $H_\mathrm{min}$, the decay of the saxion into axions instead is open and may lead to an early decay of saxions. To avoid this we have to further impose that
\begin{equation}
    \Gamma_{\text{axion}}<H_\mathrm{min}
\end{equation}
when the saxion sits on the minimum, leading to an upper bound on $\lambda$
\begin{equation}
    \lambda<\frac{4\sqrt{2}\pi f_a}{\sqrt{3}M_{\text{Pl}}}.
\end{equation}

\paragraph{Reheating during the kination era}

It is relevant to notice that if $\Gamma>H_{\text{ke}}$, reheating occurs during the kination era and the actual radiation era will start at $H_{\text{ke}}$, given in Eq.~\eqref{Hke}. The resulting temperature at this moment will be
\begin{equation}
    T_{\text{rh}}=\left(\frac{90}{\pi^2g_R}\right)^{1/4}\sqrt{H_{\text{ke}}M_{\text{Pl}}}.
\end{equation}
However, if the reheating mechanism is the decay into SM Higgs this would require $H_\text{min}>H_{\text{ke}}$ and, in terms of $\lambda$, this translates into
\begin{equation}
    \lambda>\left(\frac{3\sqrt{2}\xi^{3/2}H_I^2}{f_aM_{\text{Pl}}}\right)^{\frac{1+\alpha}{\alpha}}\frac{1}{F_{\xi}^{1/\alpha}}
\end{equation}
In our setup, this condition is forbidden by the requirements that $A<M_\mathrm{Pl}$ and that the kination era lasts less than $\sim 10$ e-folds (see Sec.~\ref{sec:parameter space}), except for a small region of parameter space when $\xi \gtrsim \mathcal{O}(100)$. We choose not to pursue this possibility any further.

\section{Parameter Space}\label{sec:parameter space}

In this section we analyze the allowed parameter space for $\lambda,\xi,H_I$.

\paragraph{Duration of the kination era}
A kination era leads to an enhancement of gravitational waves signal, that can alter the process of BBN through a change of $N_{\text{eff}}$, imposing an upper bound on the kination duration \cite{Gouttenoire:2021jhk}
\begin{equation}
    N_{\text{ke}}<11.9+\log\left(\frac{5\times10^{13}\,\text{GeV}}{H_I}\right),
\end{equation}
where $N_{\text{ke}}$ is the total number of e-folds of kination and we assumed a perfect kination era ($w=1$).
Moreover, recently it has been shown \cite{Eroncel:2025bcb} that a rolling scalar field, as can be the inflaton, leads to the growth of perturbations that behaves as radiation and can eventually come to dominate during a subsequent phase of kination. Since this bound strongly depends on the inflation model, we remain agnostic about the origin of the kination era and set a conservative bound
\begin{equation}
    N_{\text{ke}}\lesssim10.
    \label{kination bound}
\end{equation}

\paragraph{Gravitational back-reaction}
The scenario discussed so far holds as long as $\lambda$ is large enough to stop the growth of the saxion before its energy dominates that of the inflaton. In fact, if $\lambda$ is too small the saxion field comes to dominate over the inflaton, at this point kination ends and the Ricci scalar is quickly set to zero, ceasing the tachyonic growth~\cite{Figueroa:2024asq}. In order to avoid this gravitational back-reaction, one can require that at $H_{\text{max}}$ the saxion energy density is smaller than half of the inflaton energy density, which translates into
\begin{equation}
    S_{\text{max}}^2<\frac{M_{\text{Pl}}^2}{(5+4\sqrt{6\xi})\xi},
    \label{GB lowerbound}
\end{equation}
From this inequality the following lower bound on $\lambda$ is obtained:
\begin{equation}
    \lambda>\mathcal{O}(1)\left(\frac{b^2}{\pi^2\xi^{3/2}}\right)^{\frac{1}{\alpha}}\left(\frac{H_I^2(5+4\sqrt{6\xi})\xi^2}{M_\mathrm{Pl}^2}\right)^{\frac{1+\alpha}{\alpha}},
    \label{self-interaction inequality}
\end{equation}
The full dependence on $w$ is shown in Eq.~\eqref{gravitaion back reaction generic}.

If, on the contrary, gravitationa back-reaction stops the tachionic growth of the saxion, the field reaches asymptotically the value $M_{\text{Pl}}/\sqrt{\xi}$~\cite{Figueroa:2024asq} and the tachyonic growth halts, because the large field excursion of $S$ makes the term $\xi R S^2$ in the action comparable to the gravitational action $M_\mathrm{Pl}^2 R$. The field gets stuck until when its quartic potential becomes cosmologically relevant.
During this time the axion experiences the gradient produced by $V_{\cancel{U(1)}}$ and starts oscillating around the local minimum. Therefore, the axion may still have a residual kinetic energy component when the saxion begins to move, thereby also realizing kinetic misalignment.
The dynamics in this case is not covered by our discussion because the large term $\xi R S^2$ dominates the expansion of the Universe. Moreover, additional Planck-suppressed higher order operators may become relevant.

\begin{figure}
    \centering
    \includegraphics[width=1\linewidth]{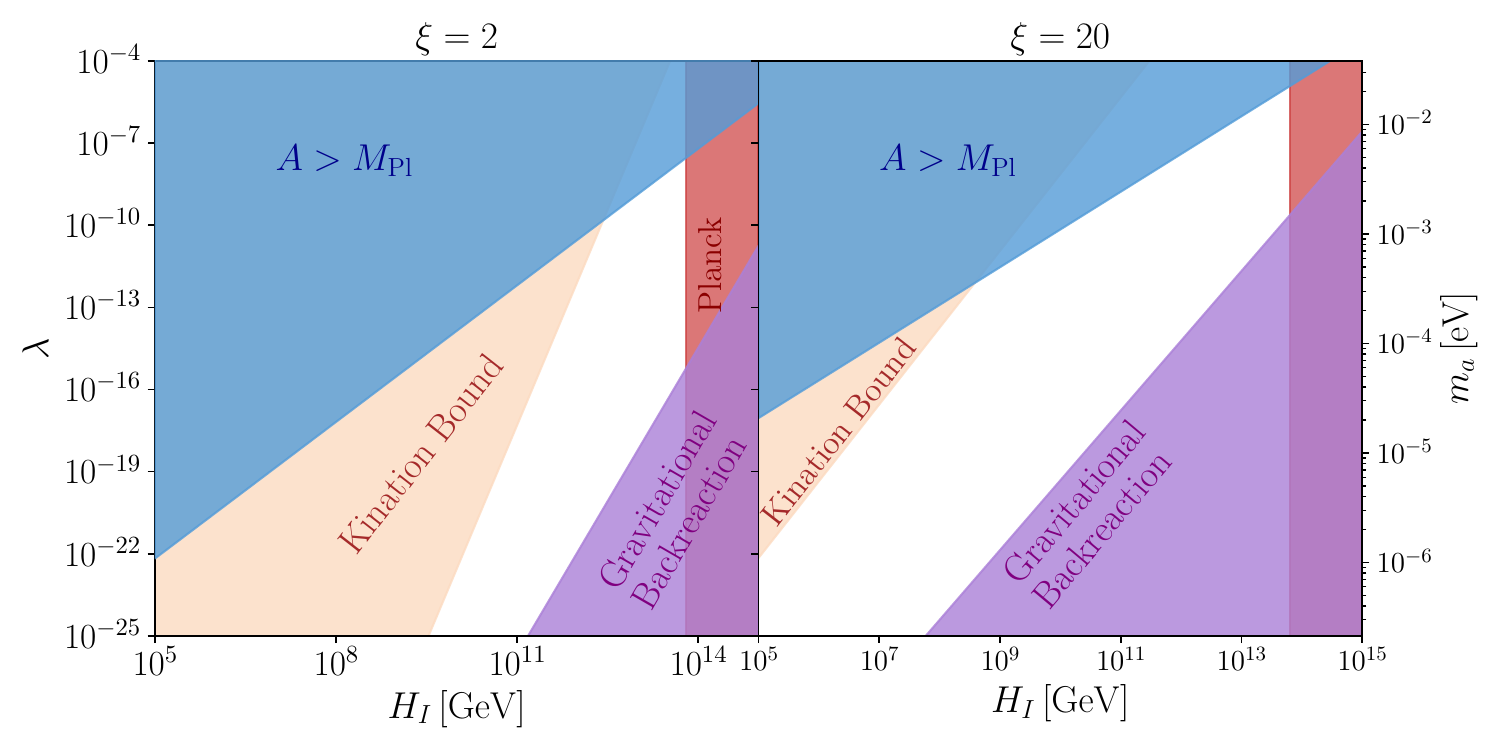}
    \caption{ In these plots we show the allowed parameter space to achieve an efficient tachyonic growth and a subsequent kick. The parameter $f_a$ is left free, $m_a$ is related to $\lambda$ by requiring the correct dark matter abundance (see Sec. \ref{sec:Axion Dark Matter Abundance}) and $A$ is fixed in order to get $\epsilon\sim1$. \textcolor[HTML]{AB80D8}{Gravitational Backreaction} refers to the saxion dominating the energy density during the tachyonic growth as computed in Eq.~\eqref{GB lowerbound}. In the region indicated with \textcolor[HTML]{fcdbc1}{Kination Bound} the duration of kination is too long violating Eq.~\eqref{kination bound}. The upper bound is provided by requiring \textcolor[HTML]{549cd8}{$A>M_{\text{Pl}}$} in order to have an efficient kick, see Eq.~\eqref{boundA}. Finally, the upper bound on $H_I$ is provided by \textcolor[HTML]{cc3e3e}{Planck} observations \cite{Planck:2018jri}. }
    \label{Parameter space}
\end{figure}

\paragraph{Higher order operator}
A strong kick in the angular direction is equivalent to requiring $\bar\epsilon\sim\mathcal{O}(1)$, for which a large value of $A$ is typically necessary (see App.~\ref{Axion kick}). We show the region where $A>M_{\text{Pl}}$ is required as a blue coloured region in our plots.

\paragraph{String decay}
If long strings form, they may reach a stable configuration known as the scaling regime \cite{Gorghetto:2018myk, Gorghetto:2020qws, Correia:2024cpk, Kim:2024wku, Kim:2024dtq, Kawasaki:2014sqa, Saikawa:2024bta, Buschmann:2021sdq, Benabou:2024msj}. During this phase, the string network radiates axions, saxions, and gravitational waves. As shown in Ref.~\cite{Gorghetto:2021fsn}, the dominant contribution is the decay into axions. Indeed, the ratio of gravitational waves versus decay into axions is suppressed by a factor of $f_a/M_{\text{Pl}}$. For the decay constants considered in this work, $10^8\,\text{GeV} < f_a < 10^{12}\,\text{GeV}$, the contribution from gravitational waves is therefore highly subdominant.
At the same time, up to 10\% of the energy can be radiated into saxions. Once produced, saxions rapidly thermalize through scatterings with the thermal bath and subsequently decay. Although this process occurs throughout the entire evolution of the string network, the entropy injected by saxion decay is expected to be negligible, as their energy remains subdominant compared to that of the thermal bath.
Therefore, the primary contribution is from axion production. Initially, axions are relativistic, but they become non-relativistic around the time when $H = m_a$, subsequently contributing to the CDM abundance. An estimate of the axion abundance originating from strings reads:~\cite{Gorghetto:2021fsn}
\begin{equation}
    h^2\Omega^{\text{string}}_a\simeq 0.1\left(\frac{\xi_\star\log_\star}{3\times10^2}\right)\left(\frac{f_a}{10^{10}\,\text{GeV}}\right)^2\left(\frac{m_a}{10^{-4}\,\text{eV}}\right)^{\frac{1}{2}}\left(\frac{10}{x_{0,a}}\right)\left(\frac{3.5}{g_*(T_\star)}\right)^{\frac{1}{4}}
\end{equation}
where $\xi_\star$ is the number of strings per Hubble patch at $H_\star=m_a$, $\log_\star=\log{(m_S/m_a)}$ and $x_{0,a}$ is the infrared cutoff of the axion emission spectrum. The presence of this extra component, combined with axions from kinetic misalignment, may lead to an overabundance. We exclude in our figures the region in which $h^2\Omega^{\text{string}}_a \gtrsim 0.1$, although in this region nothing is fundamentally wrong in the model, and the scenario can be rescued if entropy is injected in the plasma to reduce the DM abundance.

\begin{figure}[t]
    \centering
    \includegraphics[width=1\linewidth]{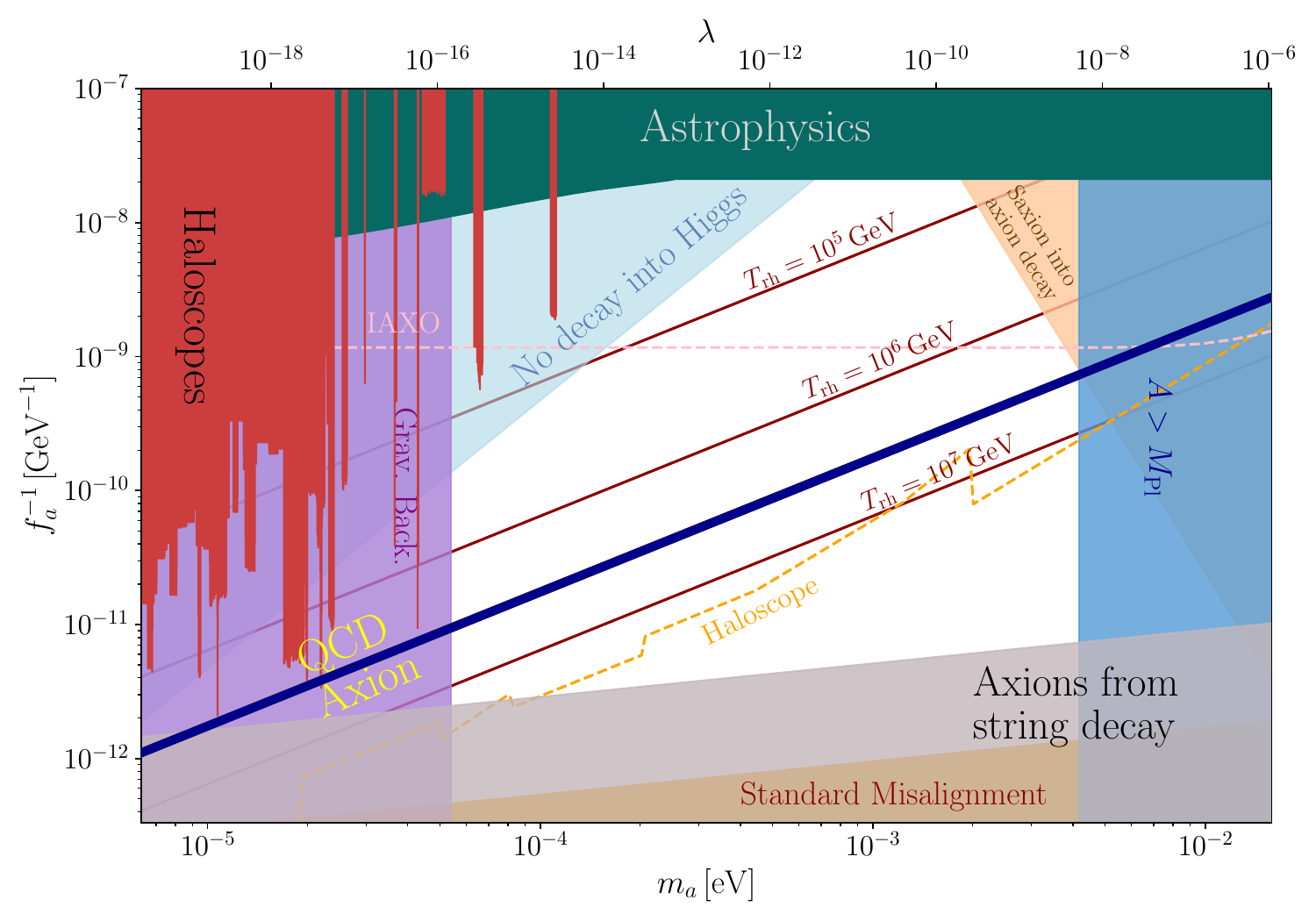}
    \caption{Parameter space $[m_a,f_a^{-1}]$ for $\xi=3, n=7, H_I=10^{13}\,\text{GeV}$. The constraints \textcolor[HTML]{076964}{Astrophysics} and \textcolor[HTML]{cc3e3e}{Haloscopes} have been taken from \cite{AxionLimits}. We show where the axion abundance is given by the  \textcolor[HTML]{c4b7bb}{string-DW decay}, for large $f_a$ dark matter is achieved via \textcolor{orange}{standard misalignment}, where the saxion mass is smaller than the Higgs mass, so reheating can not occur via \textcolor[HTML]{ADD8E6}{decay into Higgs}, conversely if the saxion mass is too large it will early \textcolor[HTML]{fec89a}{decay into axions} resulting in a cold universe, in which BBN never happened. Finally the Strong CP Problem is not solved in any point of the QCD axion line, for this reason it is colored in \textcolor[HTML]{00008B}{dark blue}. We show contours of reheating temperature, achieved via decay into Higgs bosons, Eq.~\eqref{Reheating Temperature}. With dashed lines we show projectionts of future experiments like \textcolor[HTML]{FFC0CB}{IAXO} \cite{2002clme.book.....G,Armengaud:2014gea} and a combination of \textcolor[HTML]{FFA500}{Haloscope} searches like BREAD \cite{BREAD:2021tpx}, MADMAX \cite{Beurthey:2020yuq}, ADMX \cite{ADMX:2018gho,ADMX:2018ogs,ADMX:2019uok,ADMX:2021mio,ADMX:2021nhd,Stern:2016bbw,Crisosto:2019fcj, ADMX:2025vom}.}
    \label{Parameter space 1}
\end{figure}

\begin{figure}[t]
    \centering
    \includegraphics[width=1\linewidth]{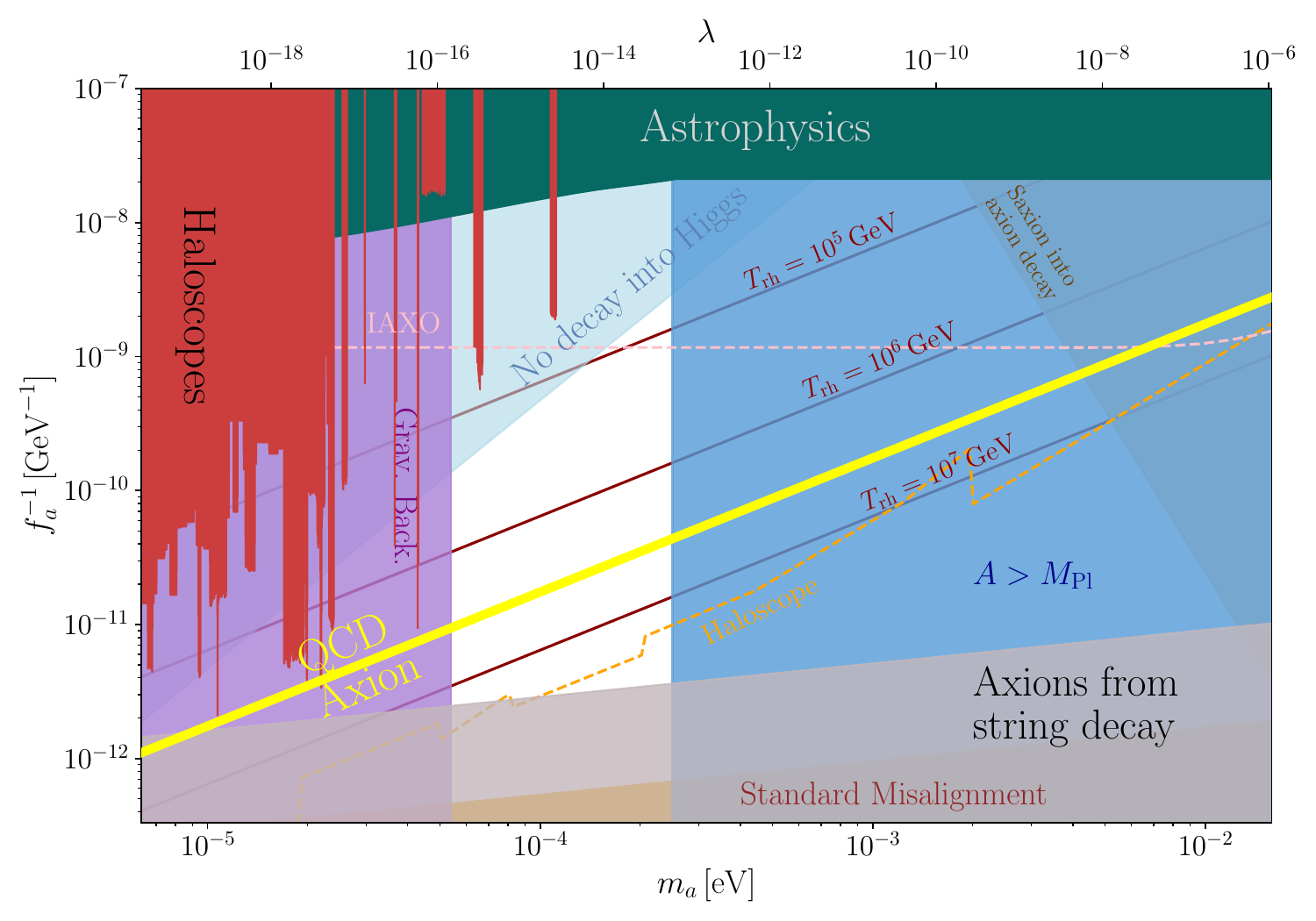}
    \caption{Same parameter space $[m_a,f_a^{-1}]$ but for $\xi=3, n=11, H_I=10^{13}\,\text{GeV}$. With this choice of parameters there's a window where the QCD axion is a viable candidate.}
    \label{Parameter space 2}
\end{figure}

\paragraph{Axion Quality Problem}
If the axion is identified with the QCD axion, an additional constraint must be taken into account. Since the QCD axion is supposed to solve the Strong CP Problem, one must ensure that the $U(1)$-violating term does not significantly shift the position of the CP-conserving minimum. Computing the contribution of $V_{\cancel{U(1)}}$ to the axion mass we have to impose that
\begin{equation}
    m_{a,\cancel{U(1)}}^2<m_a^2\theta_{\text{QCD}}<10^{-10}\,m_a^2,
\end{equation}
where $\theta_{\text{QCD}}$ is constrained by searches for the electric dipole moment of the neutron~\cite{Baker:2006ts} to be $\theta_{\text{QCD}}<10^{-10}$. We compute
\begin{equation}
     m_{a,\cancel{U(1)}}^2 = \frac{n A f_a^{n-2}}{M_\mathrm{Pl}^{n-3}}\,,
\end{equation}
and set $\theta=0$ today. Asking $A$ to reproduce $\bar\epsilon\sim\mathcal{O}(1)$, as in Eq.~\eqref{A} we get the following upper bound to the axion decay constant $f_a$:
\begin{equation}
    f_a<\left(\frac{\beta\theta_{\text{QCD}}S_{\text{max}}^{n-3}m_\pi^2 f_\pi^2}{n\sqrt{\lambda}H_{\text{max}}}\right)^{\frac{1}{n}}\,,
\end{equation}
where $m_a^2\sim m_\pi^2f_\pi^2/f_a^2$, up to $\mathcal{O}(1)$ factors. This bound is relaxed taking a larger value of $n$, however this is in tension with the bound \eqref{boundA} that constrains $A$ to be smaller than the Planck scale. We color in \textcolor[HTML]{00008B}{blue} the QCD axion line where this bound is violated.
As clear from Figs.~\ref{Parameter space 1},~\ref{Parameter space 2}, the quality problem in this scenario is a serious issue for the QCD axion. We will see in Sec.~\ref{sec:spectator saxion} that this problem is ameliorated in a scenario where reheating does not depend on saxion decays.



\section{Axion Dark Matter Abundance}\label{sec:Axion Dark Matter Abundance}

We now turn to estimate the axion DM abundance, bearing in mind that this estimate must be corrected by a detailed analysis of the string network, which goes beyond our present scope.
In each domain, the number density $n_{\theta,\text{dom}}$ generated by the kick is conserved in a comoving volume. For this reason we define the conserved quantity
\begin{equation}
    Y_{\text{dom}}=\frac{n_{\theta,{\text{dom}}}}{s},
\end{equation}
in which we are assuming that the value of the entropy density $s$ is the same in each domain. At low energies, the axion develops a periodic potential that explicitly breaks the $U(1)$ symmetry to the discrete subgroup $\theta\rightarrow\theta+2\pi/N_{\text{DW}}$. We parametrize the potential as
\begin{equation}
    V_a=m_a(T)^2f_a^2\left[1-\cos\theta\right].
\end{equation}
with $N_\mathrm{DW}=1$.
For large $\dot\theta$, the axion overshoots the barriers and keeps rotating until it gets halted at the trapping temperature $T_\text{trap}$ when
\begin{equation}
    \dot\theta(T_\text{trap})=2m_a(T_\text{trap}).
\end{equation}
For $T<T_\text{trap}$ the axion field oscillates around the minimum and starts contributing to CDM.
The charge $n_{\theta}$ is converted at $T_\text{trap}$ into the number density of axions with mass $m_a$ via the relation
\begin{equation}
    n_{a,\text{trap}}\approx 2 |n_{\theta,\text{trap}}| \,.
\end{equation}
The factor $2$ depends on the fact that, once trapped in the potential, the axion energy density does not immediately scale as $a^{-3}$, because the amplitude of oscillations is initially close to $f_a$ and the axion probes the non-harmonic region of the potential close to its maximum. The numerical result can be derived numerically~\cite{Co:2019jts} or using the action-angle formalism of classical mechanics, as done in Sec. 2 of Ref.~\cite{Eroncel:2022vjg}, to which we refer for further details.
We compute the energy density of axions today from
\begin{equation}
    \frac{\rho_{a}}{s}\approx2m_a|Y_{\theta}|.
    \label{axion energy density}
\end{equation}
As specified above, in each domain we have different $n_{\theta,\text{dom}}$ that at $T=T_\text{trap}$ are converted into axions. In order to estimate the total axion energy density we have to average over all the domains. Each domain differs from the other by a different selection of the initial angle $\theta_i$, for this reason the average on domains translates into an average over the interval $[0,2\pi)$, in perfect analogy with the post inflationary standard misalignment. Thus, we define the global $U(1)$ charge at the moment of the kick by taking the root mean square 
\begin{equation}
    \bar n_{\theta}  = \left(\frac{1}{2\pi}\int_0^{2\pi}d\theta_i\, n_{\theta,\text{dom}}^2(\theta_i)\right)^{\frac{1}{2}}=\frac{\beta AS_{\text{max}}^{n}}{\sqrt{2}M_{\text{Pl}}^{n-3}H_{\text{max}}}.
\end{equation}
As already stated in Eq.~\eqref{epsilon def}, we can write the averaged number density as
\begin{equation}
    \bar n_{\theta,\text{max}}=\bar\epsilon\, n_{S,\text{max}}\,,
\end{equation}
where $n_S$ was defined in Eq.~(\ref{epsilon def}). At this point we can compute the averaged yield at the reheating temperature
\begin{equation}
    \bar Y_{\theta}=\bar\epsilon\,\frac{n_{S,\text{max}}}{s_{\text{rh}}}\left(\frac{a_{\text{max}}}{a_{\text{rh}}}\right)^3=\bar{\epsilon}\,\frac{n_{S,\text{max}}}{s_\text{rh}}\left(\frac{H_{\text{ke}}}{H_{\text{max}}}\right)^{\frac{2}{1+w}}\left(\frac{H_{\text{rh}}}{H_{\text{ke}}}\right)^{\frac{3}{2}},
    \label{Yield general}
\end{equation}
where we assumed that reheating occurs during the scaling $\rho\propto a^{-4}$ and no further entropy injection is present. With these assumptions, expression \eqref{Yield general} simplifies to 
\begin{equation}
    \bar{Y}_\theta=140\left(\frac{100}{g_{*,\mathrm{rh}}}\right)^{\frac{1}{4}}
    \left(\frac{\bar{\epsilon}}{0.5}\right)\left(\frac{10^{-10}}{\lambda}\right)^{\frac{1}{4}}\,.
\end{equation}
It's remarkable that this quantity depends only on $\lambda$ and $\bar\epsilon$. Using the relation \eqref{axion energy density} we can relate the axion yield to $\Omega_{\text{CDM}}$, getting
\begin{equation}
    \frac{h^2\Omega_{a,0}}{h^2\Omega_{\text{CDM}}} \approx \left(\frac{m_a}{5\times 10^{-3}\,\text{eV}}\right)\left(\frac{\bar Y_{\theta}}{43}\right).
    \label{axion abundace}
\end{equation}
Fixing $\lambda$ in order that all the dark matter is given by axions produced from this mechanism we get 
\begin{equation}
    \lambda=
    1.1\times10^{-8}
    \left(\frac{100}{g_{*,\text{rh}}}\right)
    \left(\frac{\bar\epsilon}{0.5}\right)^4
    \left(\frac{m_a}{5\times 10^{-3}\,\text{eV}}\right)^4
    \,.
    \label{lambda for DM}
\end{equation}
The allowed parameter space for $[m_a,f_a^{-1}]$ is shown in Figs. \ref{Parameter space 1}, \ref{Parameter space 2}, for two different choices of $\xi$, $n$, and $H_I$.

The temperature when the axion gets trapped and starts behaving as cold dark matter is given by
\begin{align}\label{trapping temperature}
    T_\mathrm{trap} \approx & \; 1.0 \frac{g_{*S,\text{rh}}^{1/3}}{g_{*\text{rh}}^{1/4}\,g_{*S,\text{trap}}^{1/3}} \frac{f_a^{2/3} m_a^{1/3} \lambda^{1/12}}{\epsilon^{1/3}} \\
    \approx & \; 0.7\times 10^2 \,\mathrm{GeV} \left(\frac{100}{g_{*\text{rh}}}\right)^{3/4} \left(\frac{g_{*S,\text{rh}}}{g_{*S,\text{trap}}}\right)^{1/3}
    \left(\frac{f_a}{10^{10}\,\mathrm{GeV}}\right)^{2/3} \nonumber\\
    & \times \left(\frac{m_a}{5\times 10^{-3}\,\mathrm{eV}}\right)^{1/3}
    \left(\frac{\lambda}{10^{-8}}\right)^{1/12}
    \left(\frac{0.5}{\epsilon}\right)^{1/3} \\
    \approx & \; 19\,\mathrm{GeV}
    \left(\frac{\sqrt{f_a\, m_a}}{75\,\mathrm{MeV}}\right)^{4/3} \left(\frac{62}{g_{*S,\mathrm{trap}}}\right)^{1/3}
    \left(\frac{h^2\Omega_{\text{CDM}}}{h^2\Omega_{\theta,0}}\right)^{1/3}
\end{align}
where in the last line we have shown, for reference, the values relative to the QCD axion. This expression assumes, for simplicity, a temperature-independent potential.
For kinetic misalignment to be active, this temperature has to be lower than the one obtained for standard misalignent $T_\mathrm{mis}$, defined by $H(T_\mathrm{mis}) \sim m_a$. This leads to a condition
\begin{align}\label{eq:no standard misalignment}
    \frac{1}{f_a} < 4.9 \times 10^{-13} \,\mathrm{GeV}^{-1} \left(\frac{m_a}{1\,\mathrm{meV}}\right)^{1/4} \left(\frac{g_{*,\mathrm{trap}}}{62}\right)^{1/8}
\end{align}
This condition, for a constant axion potential, is less stringent than the requirement that axions from string decay do not overclose the universe, as shown in Figs.~\ref{Parameter space 1},~\ref{Parameter space 2} and~\ref{ma,fa_thermalized}. If the temperature dependence of the axion mass is taken into account, this conclusion may be modified.


\section{Saxion as a spectator field}\label{sec:spectator saxion}

\begin{figure}
    \centering
    \includegraphics[width=0.8\linewidth]{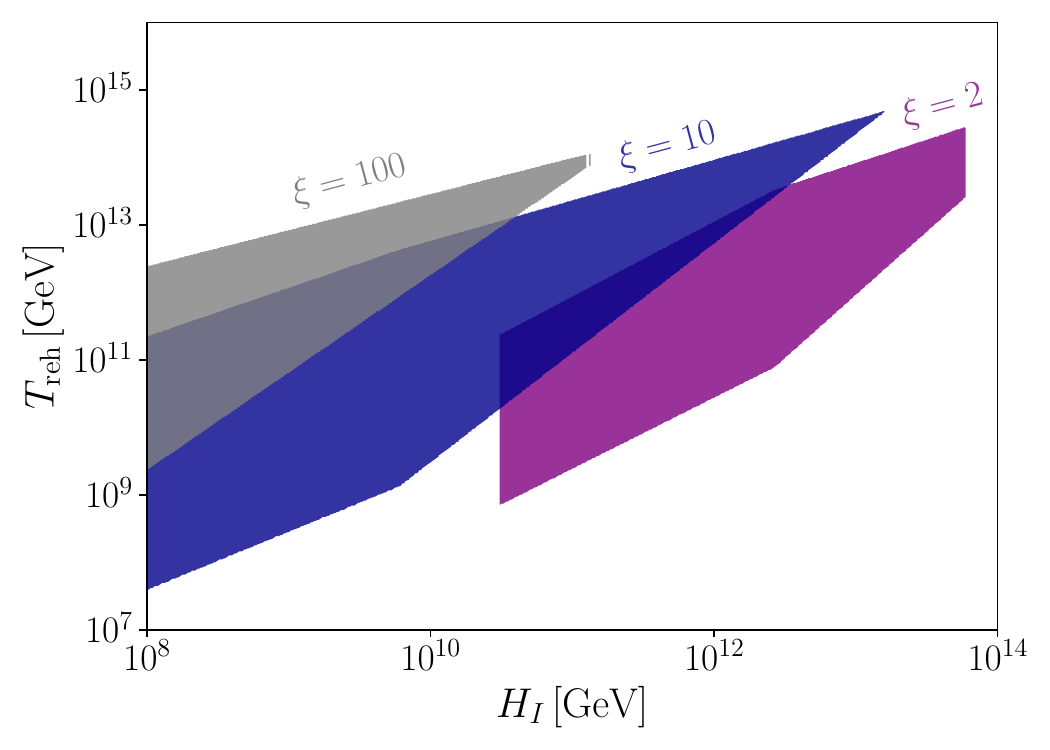}
    \includegraphics[width=0.8\linewidth]{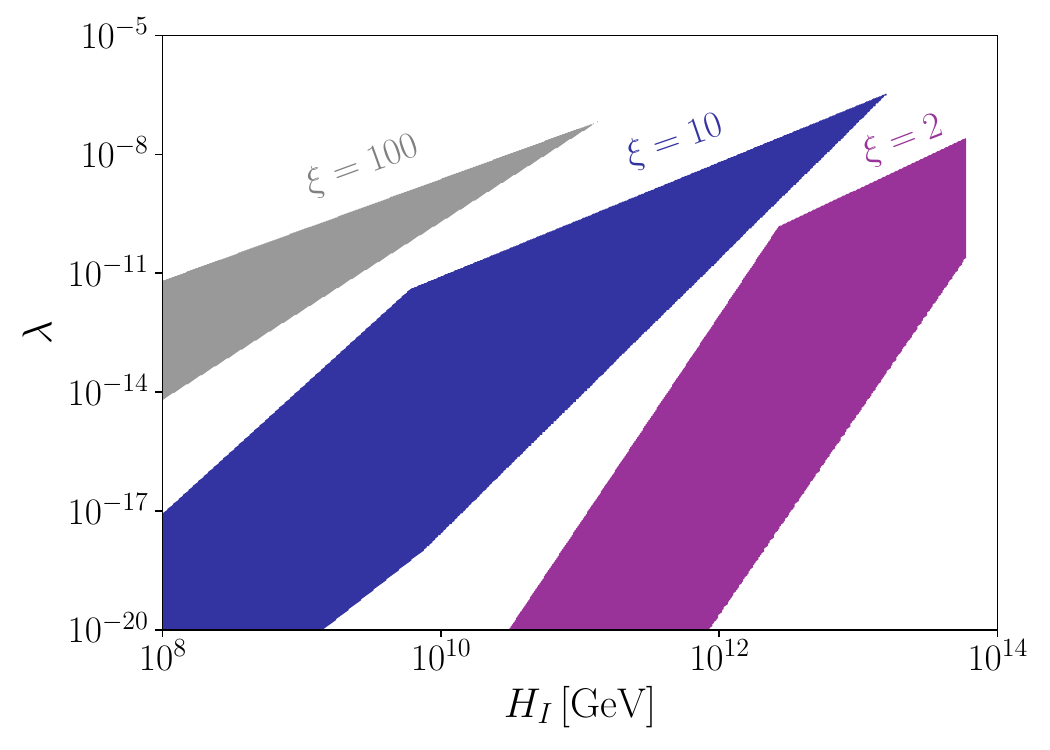}
    \caption{In these plots we show projections of the maximal allowed parameter space (colored region) in the $[T_{\text{rh}},H_I]$ plane marginalizing over $\lambda$, and in the $[\lambda,H_I]$ plane marginalizing over $T_{\text{rh}}$, for three different values of $\xi$. In this scenario the saxion energy density remains subdominant the whole time and the reheating of the universe is achieved via another unrelated mechanism.}
    \label{Second field parameter space}
\end{figure}

Here we present another scenario in which the $U(1)$ field remain subdominant during the whole cosmological history and the inflaton sector is responsible of reheating, via a mechanism which we remain agnostic about. Some natural examples can be the decay of the inflaton itself, a Ricci Reheating mechanism with the addition of a second field acting as a reheaton%
\footnote{This reheaton field may be the Higgs field itself, as shown in \cite{Opferkuch:2019zbd, Laverda:2024qjt, Laverda:2025pmg}, whithout requiring any additional BSM degree of freedom.} \cite{Opferkuch:2019zbd, Bettoni:2021zhq, Laverda:2023uqv, Figueroa:2024asq, Bettoni:2024ixe} or warm inflation \cite{Berera:1995ie}. The reheating temperature $T_{\text{rh}}$ is now a free parameter, but some constraints have to be taken into account.
In particular, in order to let the saxion experience all the tachyonic growth, kination has to end after the saxion has been stopped by the quartic barrier. For this reason, assuming that a radiation era starts right after the end of kination, we impose
\begin{equation}
    H_{\text{rh}}<H_{\text{max}}.
\end{equation}
However, if reheating occurs too late the saxion eventually comes to dominate and the actual reheating will occur at the time when saxion decays, leading to the scenario outlined above. In order to avoid this we also impose 
\begin{equation}
    H_{\text{rh}}>H_{\text{ke}}.
\end{equation}
In Fig.~\ref{Second field parameter space} we show the largest parameter space allowed for three different values of the non-minimal coupling $\xi$ and for a marginalization of $T_{\text{rh}}$ and $\lambda$. 

\subsection{Axion Dark Matter Abundance}
We compute the axion abundance in this scenario following the same procedure outlined in Sec.~\ref{sec:Axion Dark Matter Abundance}. The yield at the beginning of radiation era is given by
\begin{equation}
    Y_{\theta,\text{dom}}=\epsilon_{\text{dom}}\frac{n_{S,{\text{max}}}}{s(T_{\text{rh}})}\left(\frac{H_{\text{rh}}}{H_{\text{max}}}\right)^{\frac{2}{1+w}}, 
\end{equation}
substituting and averaging over many domains we obtain 
\begin{equation}
    \bar Y_{\theta}=\frac{9\sqrt{15}\,\bar\epsilon \,H_I^2 \xi^{3/2}}{\pi\sqrt{g_{*S,\text{rh}}}\,M_{\text{Pl}}T_{\text{rh}} F_{\xi}^{1/(1+\alpha)}\lambda^{\alpha/(1+\alpha)}}.
    \label{Yield second field}
\end{equation}
Compared to the discussion of Sec.~\ref{sec:Axion Dark Matter Abundance}, where the yield was following the relation $Y_{\theta}\propto\lambda^{-1/4}$, in this scenario the axion abundance is more sensitive to changes of $\lambda$, since $Y_{\theta}\propto\lambda^{-0.54}(\lambda^{-0.90})$ for $\xi=2(100)$, this permits a much larger parameter space as showed in Fig.~\ref{ma,fa_thermalized}.

\begin{figure}[h]
    \centering
    \includegraphics[width=1\linewidth]{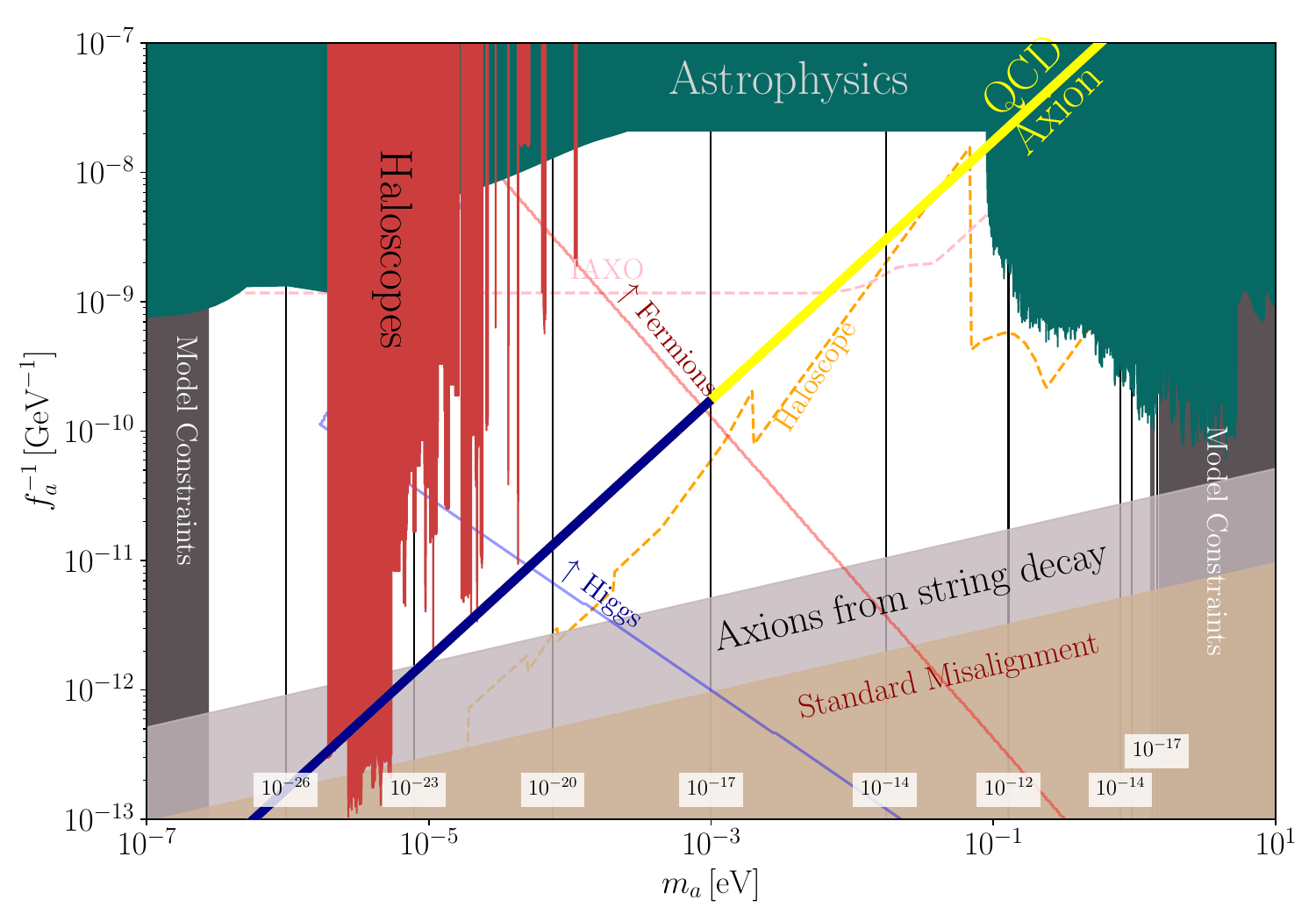}
    \caption{Maximal allowed parameter space $[m_a,f_a^{-1}]$ marginalizing on $H_I \in[10^8\,\text{GeV},10^{14}\,\text{GeV}]$, $T_{\text{rh}}\in[10^5\,\text{GeV},10^{16}\,\text{GeV}]$ (compatible with Fig.~\ref{Second field parameter space}), $y\in[10^{-10},10^{-2}]$, $\lambda_{SH}\in[10^{-20},10^{-2}]$ (values outsides of these ranges are excluded by constraints on thermalization, see Table.~\ref{Tables}) and for $\xi=5$ and $n=9$. Vertical black lines show contours of the maximum allowed value of $\lambda$, after marginalization on $H_I$ and $T_\text{rh}$. On the right side of \textcolor{blue}{blue} (\textcolor{red}{red}) line saxion damping can occur via \textcolor{blue}{Higgs} (\textcolor{red}{Fermion}) scattering. The regions dubbed ``\emph{Model Constraints}'' refer to all those bounds required to achieve an efficient kick mechanism. For a summary of the constraints considered see Table~\ref{Tables}.}
    \label{ma,fa_thermalized}
\end{figure}

\subsection{Damping saxion oscillations}

In this scenario, in which reheating already occurred by the decay of a another field, it's crucial that relic saxions thermalize and decay in order to avoid a late stage of matter dominance or an overproduction of dark matter. Unlike in the previous reheating case, discussed in Sec.~\ref{sec:rehating}-\ref{sec:Axion Dark Matter Abundance}, the saxion can now scatter with the particles in the thermal bath via a damping rate $\Gamma_{\text{damp}}$, thereby allowing the energy stored in the condensate to be damped. For consistency,  this damping mechanism has to occur after reheating, thus we require $\Gamma_{\text{damp}}<H_{\text{rh}}$. 

If the saxion reaches the minimum, its equation of state turns to be matter-like and eventually it comes to dominate over the radiation energy density. The temperature when this occurs  $T_D$ can be computed setting the ratio $\rho_S/\rho_{\text{rad}}|_{T=T_D}=1$, which for $w=1$ leads to 
\begin{equation}
    \frac{\rho_{S}(T_{\text{rh}})}{\rho_{\text{rad}}(T_{\text{rh}})}
    \left(\frac{g_{*S,\mathrm{min}}}{g_{*S,\mathrm{rh}}}\right)^{1/3}
    \frac{T_\text{min}}{T_D}
    =
    \left(\frac{g_{*S,\mathrm{min}}}{g_{*S,\mathrm{rh}}}\right)^\frac{1}{3}
    \frac{T_\mathrm{min}}{T_D}
    \frac{\rho_{S,\mathrm{max}}}{3M_\mathrm{Pl}^2}
    \frac{H_\mathrm{rh}^{\frac{8}{3(1+w)}-2}}{H_\mathrm{max}^\frac{8}{3(1+w)}}=1
\end{equation}
where $T_\mathrm{min}$ is the temperature when the saxion sits on the minimum at $S=f_a$. Solving for $T_D$ we get 
\begin{align}
    T_D & = \frac{3}{4}
    \left(\frac{g_{*S,\mathrm{min}}}{g_{*S,\mathrm{rh}}}\right)^\frac{1}{3}
    \frac{H_\text{max}^{4-\frac{8}{3 (w+1)}} H_\text{rh}^{\frac{8}{3 (w+1)}-2}}{ \lambda \, M_\text{Pl}^2}
    (3w-1)^2 \xi^2 T_\text{min}  \\
    & = 3
    \left(\frac{g_{*S,\mathrm{min}}}{g_{*S,\mathrm{rh}}}\right)^\frac{1}{3}
    \frac{\xi^2}{ \lambda}
    \frac{H_\text{max}^{8/3}}{H_\text{rh}^{2/3}}
    \frac{T_\text{min}}{M_\mathrm{Pl}^2} \\
    & =\frac{3\sqrt{15}f_a \, H_{\text{max}}^2 \, \xi^{3/2}}{\pi g_{*S,\text{rh}}^{1/2}\,M_{\text{Pl}} \, T_{\text{rh}}\sqrt{\lambda}},
\end{align}
where the $\mathcal{O}(1)$ factor contains the dependences on $g_s(T)$ and in the last equality we plugged in the temperature when $S\simeq f_a$, given by 
\begin{equation}
    T_{\text{min}}=\frac{3^{1/3}10^{1/6}}{\pi^{1/3}}\frac{g_{*S,{\text{rh}}}^{1/6}}{g_{*S,\text{min}}^{1/3}}\left(\frac{f_a}{S_{\text{max}}}\right)\left(H_{\text{max}}M_{\text{Pl}}T_{\text{rh}}\right)^{1/3}.
\end{equation}
If the saxion thermalizes after $T_D$ entropy injection occurs, and in order to avoid this we impose $T_{\text{damp}}>T_D$.

Although the saxion energy density is subdominant relative to the thermal bath, saxion decays into axions can still be dangerous. As we will show, this decay produces a population of hot axions with an abundance comparable to that of DM today, violating the bounds on the coldness of DM, and thus it must be excluded by requiring that the saxion is depleted before decaying into axions.

The number density of axions produced via saxion decay is of the same order as the number density generated by kinetic misalignment, roughly because both mechanisms consist in a transfer of energy from the saxion to the axion field. We can easily estimate the yield of axions produced from saxions decay:
\begin{equation}
    Y_{a}^{\text{decay}}=2(1-\epsilon_{\text{dom}})\frac{n_{S,\text{max}}}{s_{\text{ax}}}\left(\frac{H_{\text{rh}}}{H_{\text{max}}}\right)^{\frac{2}{1+w}}\left(\frac{H_{\text{ax}}}{H_{\text{rh}}}\right)^{3/2}\,,
\end{equation}
where quantities labelled \emph{ax} are computed at the time of saxion decay into axions.
The ratio between the axion abundance coming from kinetic misalignment and the abundance of axions produced from saxion decay is

\begin{equation}
    \frac{h^2\Omega^{\text{decay}}_{a,0}}{h^2\Omega^{\text{KMM}}_{a,0}}=\frac{Y^{\text{decay}}_{a,0}}{Y^{\text{KMM}}_{a,0}} = 
    \frac{1-\bar\epsilon}{\bar\epsilon} \frac{g_{*S,\mathrm{rh}} \, g_{*,\mathrm{ax}}^{3/4}}{g_{*S,\mathrm{ax}} \, g_{*,\mathrm{rh}}^{3/4}} \,.
\end{equation}

If the residual saxion number density decays entirely into axions, we will end up with a final population of axions produced by both mechanisms. Since we required that the average over many domains leads to $\bar\epsilon\sim0.5$, the dark matter observed today should be originated half from kinetic misalignment and half from saxion decay. Although this might not seem alarming at first glance, we must account for the fact that axions produced in saxion decays carry a large momenta, $p\simeq m_S/2$, and remain relativistic for a very long time. The temperature $T_\mathrm{nr}$ at which they finally become non-relativistic is obtained by redshifting the momentum from $T_\mathrm{ax}$ (defined by the condition $\Gamma_{S\to aa} = H_\mathrm{ax}$) to $T_\mathrm{nr}$ such that $p = m_a$.
The redshift turns out to be different if the decay takes place before or after the saxion has reached the minimum. The former option is not viable in our parameter space: we thus focus on the latter, for which $T_\mathrm{ax}< T_\mathrm{min}$.
We obtain
\begin{equation}
    T_\mathrm{nr} = \frac{g_{*S,\mathrm{ax}}^{1/3}}{g_{*S,\mathrm{nr}}^{1/3} \, g_{*,\mathrm{nr}}^{1/4}}
    \frac{\sqrt{3}\,5^{1/4}}{2 \pi} \lambda^{1/4} m_a \left(\frac{M_{\text{Pl}}}{f_a}\right)^{1/2} \,.
\end{equation}
Within the allowed parameter space, the largest possible value of $T_\mathrm{nr}$ is
\begin{equation}
    T_{\text{nr}}\simeq6.4\,\text{eV}\left(\frac{m_a}{10^{-1}\,\text{eV}}\right)
    \left(\frac{\lambda}{10^{-12}}\right)^{1/4}
    \left(\frac{10^8\,\text{GeV}}{f_a}\right)^{1/2}\,.
\end{equation}
This temperature violates bounds on coldness of dark matter requiring axions to be already non-relativistics around $T\sim10\,\text{keV}$. For this reason, we are forced to exclude this scenario and require that the saxion energy density be efficiently depleted before the decay into axions becomes efficient. Calling the temperature at this time $T_{\text{ax}}$, we have to impose $T_{\text{damp}}>T_{\text{ax}}$. It seems that this constraint has been overlooked in the existing literature regarding the kinetic misalignment mechanism.

As a last requirement, to have a consistent description of the axion trapping, saxion oscillations have to be damped before the trapping temperature Eq.~\eqref{trapping temperature}.

In summary, we impose the following constraints on the temperatures:
\begin{equation}
    T_{\text{rh}}>T_{\text{damp}}>\max(T_D,T_\mathrm{trap},T_{\text{ax}}).
\end{equation}
In the rest of this section, we will examine the damping mechanisms also exploited in Refs.~\cite{Eroncel:2024rpe, Co:2020dya, Co:2020jtv}, with rates computed in Refs.~\cite{Mukaida:2012bz, Mukaida:2012qn, Mukaida:2013xxa}.

\subsubsection{Thermalization via fermions scattering}
We start by examining the case of a saxion field coupled to fermions via the Yukawa interaction \eqref{yukawa coupling}. As shown in Sec.~\ref{sec:rehating}, the direct decay of saxions into fermions is strongly suppressed due to the large fermion mass $m_{\chi}\sim yS$ and the dominant decay channel is into axions. However, in the presence of a thermal bath, the saxion condensate scatters with the plasma, transferring energy away from the condensate and resulting in efficient damping.

If the fermions are charged under QCD, as realized in the KSVZ model, the relevant scattering rate is \cite{Eroncel:2024rpe, Mukaida:2012qn}
\begin{equation}
\Gamma_{\chi-\text{scat.}}\approx\begin{cases}
y^2\alpha_sT\quad\text{if}\quad yS<\alpha_sT,\\
y^4\frac{S^2}{\alpha_sT}\quad\text{if}\quad\alpha_sT<yS<T.
\label{fermion rate}
\end{cases}
\end{equation}
At high temperatures this process is significantly more efficient than direct decay. On the other hand, once the temperature of the universe drops below the fermion mass, a loop-induced scattering process with gluons remains operative, with rate
\begin{equation}
\Gamma_{\text{g-scat.}}\approx \frac{b\alpha_s^2T^3}{S^2},
\label{gluon rate}
\end{equation}
where we take $b=10^{-5}$ and $\alpha_s=0.1$, following Ref.~\cite{Eroncel:2024rpe}. If the reheating temperature is high enough that $yS_{\text{rh}}<\alpha_sT_{\text{rh}}$, fermions stay in thermal equilibrium throughout the period in which the saxion field remains displaced from the minimum, due to the quartic potential scaling that enforces $S/T=\text{const.}$. In this case, scatterings with fermions efficiently thermalize the saxion below the temperature
\begin{equation}
T_{\text{damp}}\simeq7\times10^8\,\text{GeV}\left(\frac{100}{g_{*,\text{damp}}}\right)^{1/2}\left(\frac{y}{10^{-4}}\right)^{2}\left(\frac{\alpha_s}{0.1}\right).
\end{equation}
However, the presence of a thermal bath to which the saxion couples also induces thermal corrections to the saxion potential. As emphasized in Refs.~\cite{Gouttenoire:2021jhk, Eroncel:2024rpe}, a large thermal mass suppresses $\epsilon$. This follows from the fact that $\epsilon$ parametrizes the competition between the terms $\dot\theta^2 S$ and $m_{S,V'}^2 S$, which can be interpreted as the centrifugal and centripetal forces stabilizing the rotation. The appearance of an additional contribution to the saxion mass thus leads to a suppression of $\epsilon$,
\begin{equation}
\epsilon\approx\mathcal{O}(1)\times\frac{m_{S,V'}^2}{m_{S,V'}^2+m_{\text{th}}^2},
\end{equation}
where $m_{S,V'}=\sqrt{V'/S}=\sqrt{\lambda}S$ is the zero-temperature mass and $m_{\text{th}}$ is the thermal correction. The latter contribution depends on the fermion mass, in particular for $T>yS$ fermions are relativistic and in equilibrium in the thermal environment, this leads to a typical quadratic potential $V_{\text{th}}\approx y^2T^2S^2$, on the other hand once the fermion population is suppressed, for $T<yS$, a thermal correction survives as a logarithmic running effect $V_{\text{th}}\approx\alpha_s^2T^4\log(y^2S^2/T^2)$ \cite{Mukaida:2012qn}. Thus we consider a thermal mass contribution given by
\begin{equation}
m_{\text{th}}=\begin{cases}
yT\quad&\text{for}\quad yS<T,\\
\alpha_{s}\frac{T^2}{S}\quad &\text{for} \quad yS>T.
\end{cases}
\end{equation}
To avoid a suppression of $\epsilon$ after reheating, we require $m_{\text{th}}\ll m_{S,V'}$.
Thermal corrections are not the only effects arising once the saxion is coupled to other fields: the same coupling generates quantum corrections, inducing an additional quartic interaction via fermion loops. To keep this contribution subdominant, we impose an upper bound on $y$, namely $y<(16\pi^2\lambda)^{1/4}$.
Moreover, a lower bound on $y$ follows from collider constraints. In our model, the Yukawa interaction \eqref{yukawa coupling} gives a late-time fermion mass $m_{\chi}=yf_a$. If fermions are too light, they could be detected at colliders through their QCD interactions; avoiding these constraints requires imposing a lower bound on their mass, which translates into $yf_a>1\,\text{TeV}$.

Once the saxion condensate thermalizes, its energy density is transferred to a thermal population of saxions kept in thermal equilibrium with the plasma by decay and inverse decay. This population survives until the temperature of the bath becomes lower than the saxion mass and the specie is Boltzmann suppressed. However, if the process that keeps the saxions in thermal equilibrium stops to be efficient, saxions freeze out and eventually decay into a hot population of axions, which will drastically affects $N_{\text{eff}}$.
In fact, the rates in Eqs.~\eqref{gluon rate} and~\eqref{fermion rate} are efficient only at high temperatures and the saxion may decouple once fermions are not anymore in the thermal bath at a temperature $T_{\text{fo}}$.
The saxion energy density is related to radiation energy via $\rho_s/\rho_{\text{rad}}\sim1/g_*(T_{\text{fo}})$, and this ratio is constant until saxions become non-relativistics at $T_{\text{nr}}$.
We compute this temperature using the redshift of the momentum. The typical saxion momentum at freeze-out is $k/a_{\text{fo}}=T_{\text{fo}}$, and it becomes non-relativistic when $k/a_{\text{nr,sax}}=m_S$. Using the fact that $g_s^{1/3}Ta=\text{const}$ we obtain
\begin{equation}
    T_{\text{nr,sax}} = m_S\left(\frac{g_s(T_{\text{fo}})}{g_s(T_{\text{nr,sax}})}\right)^{1/3}.
\end{equation}
Calling $T_{\text{ax}}$ the temperature when the saxion decay into axion, we can compute the saxion energy density at this time getting
\begin{equation}
    \rho_S(T_{\text{ax}})=\frac{\rho_{\text{rad}}(T_{\text{nr,sax}})}{g_*(T_{\text{nr,sax}})}
    \left(\frac{g_{*S}(T_{\text{nr,sax}})}{g_{*S}(T_{\text{fo}})}\right)^{4/3}
    \left(\frac{g_{*S}(T_{\text{ax}})}{g_{*S}(T_{\text{nr,sax}})}\right)
    \left(\frac{T_{\text{ax}}}{T_{\text{nr,sax}}}\right)^3\,.
    \label{energy at Taxion}
\end{equation}
This energy density is converted into axions behaving as dark radiation $\rho_a = \frac{7}{8}\left(\frac{4}{11}\right)^{4/3}\Delta N_\mathrm{eff} \, \rho_\gamma$. Converting $\rho_S$ into $\rho_a$, from Eq.~\eqref{energy at Taxion} we get
\begin{equation}
    \Delta N_{\text{eff}}=
    \left(\frac{11}{4}\right)^{4/3} \frac{8}{7} \frac{2^{9/4}\pi}{3^{1/2}5^{1/4}}
    \frac{g_{*S}(T)^{4/3} \, g_{*,\text{ax}}^{1/4}}{g_{*S,\text{fo}} \, g_{*S,\text{ax}}^{1/3}}
    \frac{f_a}{\sqrt{m_S \,M_{\text{Pl}}}} \,.
\end{equation}
The strongest bound on $\Delta N_{\text{eff}}$ comes from the CMB \cite{ParticleDataGroup:2020ssz} which sets $\Delta N_{\text{eff}} \lesssim 0.3$. Setting this bound at the time of CMB, translates into imposing
\begin{equation}
    m_S>\mathcal{O}(1)\times\frac{f_a^2}{M_{_{\text{Pl}}}}.
\end{equation}
In Fig.~\ref{ma,fa_thermalized} we show the region where all these constraints are satisfied and damping via fermions is viable.

\subsubsection{Thermalization via Higgs}
We now move on to consider the case in which a coupling to the Higgs boson is responsible for thermalizing the saxion. In addition to the perturbative decay, Eq.~\eqref{Higgs decay rate}, thanks to presence of a thermal bath, further scattering processes can make the damping mechanism even more efficient. A first process to be considered is the scattering $SH\rightarrow SH$, which transfers energy from the saxion condensate to saxion particles in thermal equilibrium, with the following rate~\cite{Co:2020dya}
\begin{equation}
    \Gamma_{SH\rightarrow SH}\approx \frac{\lambda_{SH}^2T^3}{6\pi m_{H,\text{th}}^2}\quad \text{for} \quad m_S\ll\alpha_2T\quad\text{and}\quad m_H\ll y_t T,
\end{equation}
where the zero temperature Higgs mass is given by $m_H^2=\lambda_{\text{SH}}(S^2-f_a^2)+m_{H,0}^2$, while the thermal mass is $m_{H,\text{th}}^2=y_tT^2$ and we set $\alpha_2=1/30$ and $y_t\sim1$.
Another relevant process is the scattering $SH\rightarrow HZ$ which leads to the decay rate \cite{Co:2020dya}
\begin{equation}
    \Gamma_{SH\rightarrow HZ}=\alpha_2\frac{\lambda_{SH}^2S^2}{T}\quad \text{for} \quad m_H\ll y_t T.
\end{equation}
What we find is that the latter process is the most efficient and leads to thermalization in a wide region of the parameter space. 

Thermal corrections apply also in this case. In particular, a quadratic thermal potential  $V\approx\lambda_{SH}T^2S^2$ arises for the case $T>m_H$ and a logarithmic potential $V\approx\alpha_2^2T^4\log(m_H^2/T^2)$ for $T<m_H$. This contribution leads to a thermal mass
\begin{equation}
    m_{\text{th}}=\begin{cases}
        \sqrt{\lambda_{SH}}T\quad&\text{for}\quad \sqrt{\lambda_{SH}}S<T,\\
        \alpha_2\frac{\sqrt{\lambda_{SH}}T^2}{\sqrt{\lambda_{SH}S^2+m_{H}^2}}\quad &\text{for} \quad \sqrt{\lambda_{SH}}S>T,
    \end{cases}
\end{equation}
that we impose to be much smaller than the zero temperature mass $m_S$ in order to avoid the suppression of $\epsilon$ described above. Moreover, quantum corrections coming from closing an Higgs loops have to be kept under control, for this reason we impose $\lambda_{SH}<(16\pi\lambda)^{1/4}$.

Thermalization via Higgs keeps the saxion in thermal equilibrium even after it becomes non relativistic. Indeed, the Higgs coupling introduces a mixing angle $S-H$ given by
\begin{equation}
    \theta_{SH}\simeq-2\lambda_{SH}\frac{f_av_{\text{EW}}}{m_h^2-m_S^2}.
\end{equation}
that efficiently couples the saxion to the Standard Model. An example can be the interaction with SM leptons via the decay
\begin{equation}
    \Gamma_{Sff}=\frac{1}{8\pi}\theta_{SH}^2y_f^2 m_S.
\end{equation}
This interaction is efficient at low temperatures and keeps the saxion in equilibrium until $T\sim m_S$, when it is Boltzmann suppressed and no constraints on possible saxion relics have to be taken into account.
However, as pointed out in Ref.~\cite{Co:2020dya}, if the last decay occurs too late it may alter $N_{\text{eff}}$. In particular, if saxions decays after neutrino decoupling, the injection of energy into the thermal bath makes neutrinos relatively cooler, resulting into a reduction of $N_{\text{eff}}$. In order to avoid this, we follow Ref.~\cite{Co:2020dya} and ask that the saxion abundance is sufficiently Boltzmann suppressed before neutrino decoupling, imposing
\begin{equation}
    m_S>4\,\text{MeV}.
\end{equation}

Taking into account all these constraints, we plot the maximal allowed parameter space in Fig.~\ref{ma,fa_thermalized}, where we performed a marginalization over the parameters $H_I$, $T_{\text{rh}}$, the Yukawa coupling $y$, the coupling with the Higgs $\lambda_{SH}$, and we set $\lambda$ in order to reproduce the observed dark matter via the relation in Eq.~\eqref{Yield second field}.
Compared to the scenario in Sec.~\ref{sec:Axion Dark Matter Abundance}, we see that the quality problem for the QCD axion is less severe, and larger values of $n$ are allowed.

\begin{table}[h]
\centering
\begin{tabularx}{\textwidth}{|c|X|}
\hline
\textbf{Model Constraints}&\\
\hline
$H_{\text{rh}}<H_{\text{max}}$& Reheating has to occur after the saxion reached the maximum in order to not affect the kick mechanism.\\
\hline
$H_{\text{rh}}>H_{\text{ke}}$& Reheating has to occur before the saxion dominates the energy density, otherwise we reach the scenario outlined in section \ref{sec:rehating}.\\
\hline
$A<M_{\text{Pl}}$ & We exclude the region where a super-planckian coupling is required, see around Eq.~\eqref{boundA}.\\
\hline
$S_{\text{max}}^2<\frac{M_{\text{Pl}}^2}{(5+4\sqrt{6\xi})\xi}$ & The saxion has to be subdominat during the tachyonic growth and no gravitational backreaction must occur.\\
\hline
$\Delta N_{\text{rh}}<10$ & The duration of a kination era has an upper bound, Eq.~\eqref{kination bound}.\\
\hline
$T_{\text{trap}}>T_{\text{BNN}}$ & We impose the $U(1)$ charge to be converted into axions before the beginning of BBN.\\
\hline
\end{tabularx}\vspace{0.1cm}
\centering
\begin{tabularx}{\textwidth}{|c|X|}
\hline
\textbf{Fermions} & \\
\hline
$T_{\text{damp}}>\text{max}(T_\text{trap},T_D,T_{\text{axion}})$& Damping before the trapping and the saxion domination\\
\hline
$T_{\text{damp}}<T_{\text{rh}}$& We require a thermal bath to scatter with  \\
\hline
$m_{S,V'}^2\gg m_{\text{th}}^2$& Thermal corrections  to the saxion potential have to be subdominant\\
\hline
$y<(16\pi^2\lambda)^{1/4}$ & Quantum corrections to the saxion potential have to be subdominant\\
\hline
$yf_a>1\text{Tev}$ & In order to avoid collider constraints we impose a large fermion mass\\
\hline
$m_S>\mathcal{O}(1)\times\frac{f_a^2}{M_{\text{Pl}}}$ & In order to not have a large contribution to $\Delta N_{\text{eff}}$ form the late decay of saxions into axions.\\ 
\hline
\end{tabularx}\vspace{0.1cm}
\centering
\begin{tabularx}{\textwidth}{|c|X|}
\hline
\textbf{Higgs} & \\
\hline
$T_{\text{damp}}>\text{max}(T_\text{trap},T_D,T_{\text{axion}})$& Damping before trapping, saxion domination and decay into axions\\
\hline
$T_{\text{damp}}<T_{\text{rh}}$& We require a thermal bath to scatter with  \\
\hline
$m_{S,V'}^2\gg m_{\text{th}}^2$& Thermal corrections  to the saxion potential have to be subdominant\\
\hline
$\lambda_{SH}<(16\pi^2\lambda)^{1/4}$ & Quantum corrections to the saxion potential have to be subdominant\\
\hline
$m_S>4\,\text{MeV}$ & Saxions have to be Boltzmann suppressed before neutrino decoupling.\\ 
\hline
\end{tabularx}
\caption{Summary of constraints used for the marginalization in the scenario presented in Sec.~\ref{sec:spectator saxion}, in which reheating is not due to saxion decays.}\label{Tables}
\end{table}

\section{Conclusion}\label{sec:conclusions}

In this work, we studied the kinetic misaligment mechanism in a post-inflationary setup. The new ingredient is a non-minimal coupling of the complex scalar field to gravity, that generates a negative mass term during a stiff era at the end of inflation. The field thus develops a large vev, and start rotating in the complex field space thanks to $U(1)$ breaking higher dimensional operators.
The rotation of the field is sustained by the conservation of the generated $U(1)$ charge, with the symmetry being restored because the radial mode redshifts and the higher-order operator becomes negligible.
Due to the randomization of the axion field in our universe after the spontaneous symmetry breaking, this kick mechanism leads to the formation of distinct regions in which different $U(1)$ charges are generated.
Their evolution proceeds independently until the axion potential breaks the continuous symmetry to a discrete one, converting the  $U(1)$ charge into an axion number density and, consequently, dark matter.
The evolution of these domains requires a non-linear analysis which we postpone to future work. In this paper, we estimated the axion dark matter abundance taking an average on different domains of the $U(1)$ charge produced by the kick.

A crucial ingredient is the conversion of the energy in the radial mode into SM radiation. In a first scenario, the decay of the saxion into Higgs bosons reheats the Universe initiating the hot big bang history. A second scenario is possible, in which the saxion does not dominate the energy budget of the universe and another reheating mechanism operates to produce the SM plasma. In this case we showed that, over a sizeable portion of parameter space, the saxion energy density is efficiently damped via scattering with charged fermions or with the Higgs boson.

This implementation of kinetic misalignment gives rise to a rich phenomenology. The presence of topological defects, particularly the formation of cosmic strings, is characteristic of this setup and can modify the field evolution; this will be carefully studied with lattice simulations in future works. At the same time, our scenario is free from the isocurvature constraints that affect the pre-inflationary case.
Finally, the allowed ranges of axion mass and decay constant displayed in Figs.~\ref{Parameter space 1}, \ref{Parameter space 2} and ~\ref{ma,fa_thermalized} is testable in near future axion experiments, thus making our model particularly interesting.

\section*{Acknowledgment}
We thank Marco Gorghetto, Adriana Menkara, Pedro Schwaller, G\'eraldine Servant for useful discussions.
E.M. is supported by the Italian Ministry for University and Research (MUR) Rita Levi-Montalcini grant ``New directions in axion cosmology'', and acknowledges additional support by Istituto Nazionale di Fisica Nucleare (INFN) through the Theoretical Astroparticle Physics (TAsP) project.
The work of R.N. is supported by the Deutsche Forschungsgemeinschaft under Germany’s Excellence Strategy - EXC 2121 Quantum Universe - 390833306.
\clearpage

\appendix 
\section{Non-minimally coupled $\text{U}(1)$ field}\label{NMC}

\subsection{Action and energy-momentum tensor}

In this appendix, we review some features of complex scalar fields non-minimally coupled to gravity. Writing down the action in Jordan frame
\begin{equation}
    S=\int \text{d}^4x\sqrt{-g}\left[\frac{M_{\text{Pl}}^2}{2}R-g^{\mu\nu}\partial_\mu\Phi\partial_\nu\Phi^{\dagger}-\xi R|\Phi|^2-V(\Phi,\Phi^{\dagger})\right]\,,
    \label{action model appendix}
\end{equation}
we can compute the stress energy tensor from the definition 
\begin{equation}
    T_{\mu\nu}= \frac{-2}{\sqrt{-g}}\frac{\delta S_{\Phi}}{\delta g^{\mu\nu}}\,.
\end{equation}
This leads to the expression 
\begin{align}
\label{Tmunu_complex}
T_{\mu\nu} = & \underbrace{\partial_{\mu}\Phi^*\,\partial_{\nu}\Phi+\partial_{\nu}\Phi^*\,\partial_{\mu}\Phi-g_{\mu\nu}\,\left(g^{\alpha\beta}\,\partial_{\alpha}\Phi^*\,\partial_{\beta}\Phi
+V(\Phi,\Phi^*)
\right)}_{\substack{\text{canonical term}}} \nonumber \\
& + \underbrace{2\xi\Bigl(G_{\mu\nu}+g_{\mu\nu}\nabla^{\sigma}\nabla_{\sigma} -\nabla_{\mu}\nabla_{\nu}\Bigr)\,|\Phi|^2}_{\substack{\text{non-minimal coupling}}}\,,
\end{align}
in which we remark the presence of an extra term, containing geometrical contributions.
Applying the condition of homogeneity we can interpret our a field as a perfect fluid with $\rho_{\Phi}=-g^{00}T_{00}$ and $p_{\Phi}=g^{ij}T_{ij}/3$, explicitly
\begin{subequations}
\begin{align}
    \rho_{\Phi}&=\dot\Phi\dot\Phi^{*} +\frac{1}{a^2}|\vec\nabla\Phi|^2 +V+6\xi\left(H^2|\Phi|^2+H(\dot\Phi\Phi^*+\Phi\dot\Phi^*) - \frac{1}{3a^2}\nabla^2|\Phi|^2\right), \label{NMC energy}\\
    p_{\Phi}&=\dot\Phi\dot\Phi^{*} -\frac{1}{3a^2}|\vec\nabla\Phi|^2 - V + \nonumber \\
    & + 2\xi\left[
    -\left(H^2+2\frac{\ddot a}{a}\right)|\Phi|^2 - (\ddot\Phi\Phi^* + \Phi\ddot\Phi^* +2|\dot\Phi|^2) - 2H\left(\dot\Phi\Phi^*+\Phi\dot\Phi^*\right) + \frac{2}{3a^2}\nabla^2|\Phi|^2\right].\label{NMC pressure}
\end{align}
\end{subequations}

\subsection{Power spectrum at the end of inflation}

During inflation,  we can assume an equation of state with $w\simeq-1$, leading to a Ricci scalar $R=12H^2_{\text{inf}}$. During this era $\Phi$ is heavy with $m_{\text{eff}}\sim H_{\text{inf}}$, therefore the field is stuck around the the minimum $\Phi=0$. Here, we compute the power spectrum for the field perturbations at the end of inflation. We start moving to conformal time $d\tau=dt / a$ and $\Phi=\sigma/a$. Neglecting the potential $V$, the action reads
\begin{equation}
    S_{\Phi}=\int d\tau d^3x\left(|\sigma'|^2-|\nabla\sigma|^2+a^2\left(\frac{1}{6}-\xi\right)R|\sigma|^2\right),
    \label{sigma action}
\end{equation}
we promote $\sigma$ to a quantum operator
\begin{align}
    \hat{\sigma}(\tau,x)&=\int\frac{d^3k}{(2\pi)^3}\left(\sigma_k(\tau)\hat{a}_ke^{-ikx}+\sigma^*_k(\tau)\hat{b}^{\dagger}_ke^{ikx}\right),\\
    \hat{\sigma}^{\dagger}(\tau,x)&=\int\frac{d^3k}{(2\pi)^3}\left(\sigma_k(\tau)\hat{b}_ke^{-ikx}+\sigma^*_k(\tau)\hat{a}^{\dagger}_ke^{ikx}\right),
\end{align}
and from \eqref{sigma action} we find $\sigma_k$ to solve the equation of motion
\begin{equation}
    \sigma''_k+\left(k^2-\frac{\nu^2-1/4}{\tau^2}\right)\sigma_k=0,
    \label{Sasaki-Mukanov}
\end{equation}
where we used $R=12H^2=12/(a^2\tau^2)$ and, as usual we defined 
\begin{equation}
    \nu^2=\frac{1}{4} - 12\left(\xi - \frac16\right) < 0 \,.
    \label{nu}
\end{equation}
Equation~\eqref{Sasaki-Mukanov} has the following solution
\begin{equation}
    \sigma_k(\tau)=\sqrt{-\tau}\left[\alpha_kH_{\nu}^{(1)}(-k\tau)+\beta_kH_{\nu}^{(2)}(-k\tau)\right]\,.
\end{equation}
which is valid for complex $\nu$, with $\tau < 0$ during inflation.
The asymptotic behaviour of Hankel functions, with $\nu=i\mu$ and $z=-k\,\tau$, is
\begin{align}
    H_{i\mu}^{(1)}(z) &\sim \sqrt{\frac{2}{\pi z}} e^{\frac{\mu\pi}{2}} e^{i \left( z - \frac{\pi}{4} \right)},\\
    H_{i\mu}^{(2)}(z) &\sim \sqrt{\frac{2}{\pi z}} e^{-\frac{\mu\pi}{2}} e^{-i \left( z - \frac{\pi}{4} \right)}\,.
\end{align}
We require that the solution matches the Bunch-Davies vacuum for modes well inside the horizon
\begin{equation}
    \sigma_k(k\tau\rightarrow-\infty)\approx\frac{e^{-ik\tau}}{\sqrt{2k}},
\end{equation}
thus setting
\begin{equation}
\alpha_k=\frac{\sqrt{\pi}}{2}e^{\frac{\pi}{4}(i-2\mu)}\,, \qquad \beta_k=0 \,.
\end{equation}
We get then
\begin{equation}
    \sigma_k(\tau)=\sqrt{-\frac{\pi\tau}{4}}e^{(i-2\mu)\frac{\pi}{4}}H^{(1)}_{i\mu}(-k\tau).
\end{equation}
At this point we can compute the power spectrum of $\Phi$ and $\Phi'$
\begin{align}
    \langle0|\hat{\Phi}\hat{\Phi^{\dagger}}|0\rangle&=\int\frac{dk}{k}\Delta^2_{\Phi}(k,\tau),\\
    \langle0|\hat{\Phi}'\hat{\Phi}'^{\dagger}|0\rangle&=\int\frac{dk}{k}\Delta^2_{\Phi'}(k,\tau),
\end{align}
getting
\begin{align}
    \Delta_{\Phi}^2(k,\tau)&=\frac{k^3}{2\pi^2a^2}|\sigma_k|^2,\\
    \Delta_{\Phi'}^2(k,\tau)&=\frac{k^3}{2\pi^2a^2}|\sigma_k'-aH\sigma_k|^2.
\end{align}
Our field is heavy during inflation, and from definition \eqref{nu} is imaginary $\nu=i\mu$ with
\begin{equation}
    \mu^2=12\left(\xi-\frac{1}{6}\right)-\frac{1}{4} \approx 12\xi \,.
\end{equation}
For super-horizon modes $-k\tau<1$ and for $\mu\gg1$ the Hankel function can be approximated as~\cite{Dunster1990Hankel}
\begin{equation}
    H_{i\mu}^{(1)} (x) \approx e^\frac{\mu\pi}{2}\sqrt{\frac{2}{\mu\pi}} \exp\{i(\mu\log\frac{x}{2}-\phi_{\mu,0})\}
\end{equation}
with $\phi_{\mu,s} = \arg \Gamma(1+s+i\mu)$.
The power spectrum we get is therefore
\begin{align}
    \Delta_{\Phi}^2(k,\tau) & \approx \frac{H^2_I}{4\pi^2 \mu}(-k\tau)^3\\
    \Delta_{\Phi'}^2(k,\tau)&\approx a^2\frac{H^4_I}{4\pi^2}\mu\,(-k\tau)^3.
\end{align}


At the end of inflation, the inflaton acquires a large kinetic energy, transitioning its equation of state from a quasi-cosmological constant-like regime ($w=-1$) to a stiff one ($w>1/3$). This transition causes a sign change in the Ricci scalar $R=(1-3w)H^2$, thereby providing a tachyonic mass to the $\Phi$ field. From now on we are interested in studying the dynamics of the angular mode, for this reason we change basis in field space to $\Phi=S/\sqrt{2}e^{i\theta}$. To set the initial conditions for the subsequent dynamics, we take the RMS of the power spectrum integrated over superhorizon modes 
\begin{align}
    S_I^2 & = \langle \hat{S}^2\rangle=2\langle \hat{\Phi}\hat{\Phi^{\dagger}}\rangle=2\int_0^1\frac{dk}{k}\Delta_{\Phi}^2(k,\tau_I) = \frac{H_I^2}{6\pi^2(12\xi)^{1/2}}\\
    \dot S_I^2 & = \frac{1}{a^2} \langle \hat{S'}^2\rangle \simeq \frac{1}{a^2} \langle \hat{\Phi}'\hat{\Phi'^{\dagger}}\rangle=\int_0^1\frac{dk}{k}\Delta_{\Phi'}^2(k,\tau_I) = \frac{H_I^2}{12\pi^2}(12\xi)^{1/2}
\end{align}
in wich we assumed $\langle S'^2\rangle\simeq\langle\theta'^2S^2\rangle$. 

\section{Tachyonic growth of the saxion}\label{sec:appendix tachyonic growth}
In this Appendix we give more details on the saxion behavior during its tachyonic growth. The saxion potential around $S=0$ is given by
\begin{equation}
    V(S) = \frac{1}{4}\lambda f^4+\frac{1}{2}(\xi R-\lambda f^2)S^2+\frac{\lambda}{4}S^4\,.
    \label{saxion potential}
\end{equation}
Right after the transition from inflation to kination the Ricci scalar turns negative providing a new minimum
\begin{equation}\label{Smin}
    S_{\text{min}}=\sqrt{f_a^2+|R|\frac{\xi}{\lambda}}= H\sqrt{\frac{f_a^2}{H^2}+3(3w-1)\frac{\xi}{\lambda}} \simeq H\sqrt{\frac{3(3w-1)\xi}{\lambda}}\,.
\end{equation}
The position of this minimum is clearly time dependent and, after some efolds, it settles back to the minimum today $S_{\text{min}}^0=f_a$. In a stiff era with constant $1/3<w<1$, we have 
\begin{equation}
    S_{\text{min}}\propto a^{-2}-a^{-3},
\end{equation}
and the minimum is restored in a number of e-folds after inflation
\begin{equation}
    N_{\text{rest}}=\frac{2}{3(1+w)}\log{\left(\frac{H_I}{f_a}\sqrt{\frac{3(3w-1)\xi}{\lambda}}\right)}.
\end{equation}
The value of the potential in the minimum is 
\begin{equation}
    V(S_{\text{min}})\simeq-\frac{9(3w-1)^2\xi^2}{4\lambda}H^4\propto a^{-8}-a^{-12}\quad \text{for}\quad 1/3<w<1.
\end{equation}
\paragraph{Analytical solution during tachyonic regime}
During the tachyonic regime the equation of motion is well approximated by
\begin{equation}
    \ddot S+3H\dot S+\xi RS\simeq0,
    \label{initial equation}
\end{equation}
assuming the ansatz 
\begin{equation}
    S(t)=bS_I\left(\frac{H_I}{H(t)}\right)^\alpha,
    \label{growing solution appendix}
\end{equation}
we solve for $\alpha$ getting
\begin{equation}
    \alpha=\frac{-(1-w)+\sqrt{(1-w)^2+\frac{16}{3}\xi(3w-1)}}{2(1+w)}.
\end{equation}
which grows monotonically from $0$ to $\sqrt{2\xi/3}$ for $1/3\leq w \leq 1$ and $\xi\geq 1/6$.
We notice that the full numerical solution has a short initial transient during which it grows faster than \eqref{growing solution appendix} with $b=1$, which results in a mismatch of order one between analytical and numerical solutions. This mismatch can be solved by taking~\cite{Opferkuch:2019zbd, Figueroa:2024asq}
\begin{equation}
    b=\frac{1}{2}+\frac{1}{\sqrt{3w-1}} \,.
\end{equation}
Inside the text we adopted for simplicity $w=1$, so $b=1$ is a good approximation, however for $w\simeq1/3$ this correction becomes relevant.

\paragraph{Energy density evolution}
The saxion equation of state during the tachyonic growth can be computed using Eqs.~\eqref{NMC energy} and~\eqref{NMC pressure}. Specifying them for the radial mode $S$ and neglecting the gradients, they read
\begin{subequations}
\begin{equation}\label{saxion energy density}
   \rho_S=\frac{1}{2}\dot S^2+V+3\xi(H^2S^2+2H\dot S S),
\end{equation}
\begin{equation}
   p_S=\frac{1}{2}\dot S^2-V+\frac{\xi}{3}(3H^2-R)S^2-2\xi (\dot S^2+ S\ddot S+2HS\dot S).
\end{equation}
\end{subequations}
We can now compute $w_S=p_S/\rho_S$ at the beginning of the tachyonic growth. Since the position of the minimum is set by the term of non-minimal coupling, we can neglect the quartic potential. Thus, using Eq.~\eqref{initial equation} we plug $\ddot S=-3H\dot S-\xi RS$ and, from the solution \eqref{growing solution appendix}, $\dot S=3(1+w)/2\alpha S H$, getting
\begin{equation}
    w_S=\frac{3(1+w)-\sqrt{9(w-1)^2+48\xi(3w-1)}}{6}.
\end{equation}
For $w=1/3$ the dependence on $\xi$ cancel out and we get $w_S=1/3$, so the saxion behaves like radiation since $\alpha=0$, however for $w>1/3$ the tachyonic growth can compensate the energy dilution due to expansion and for some values of $\xi$ the saxion energy density can actual growth. To estimate the $\xi_{\text{crit}}$ that makes the saxion behaving as a cosmological constant, $w_S=-1$, we follow \cite{Opferkuch:2019zbd} and find
\begin{equation}
    \xi_{\text{crit}}=\frac{3(1+w)}{2(3w-1)}.
\end{equation}
So, for $w=1$ we get that for $\xi>3/2$ the saxion energy density will actually growth despite the expansion. 

\begin{figure}[t]
    \includegraphics[width=0.5\linewidth]{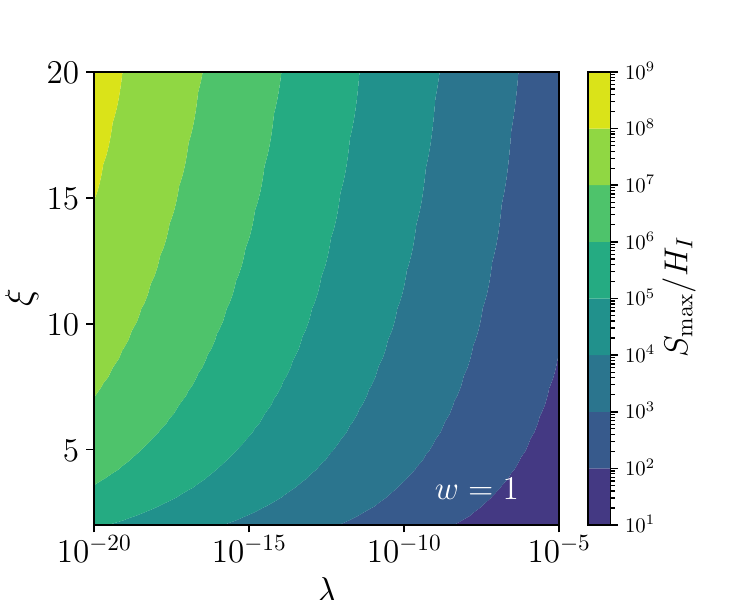}
    \includegraphics[width=0.5\linewidth]{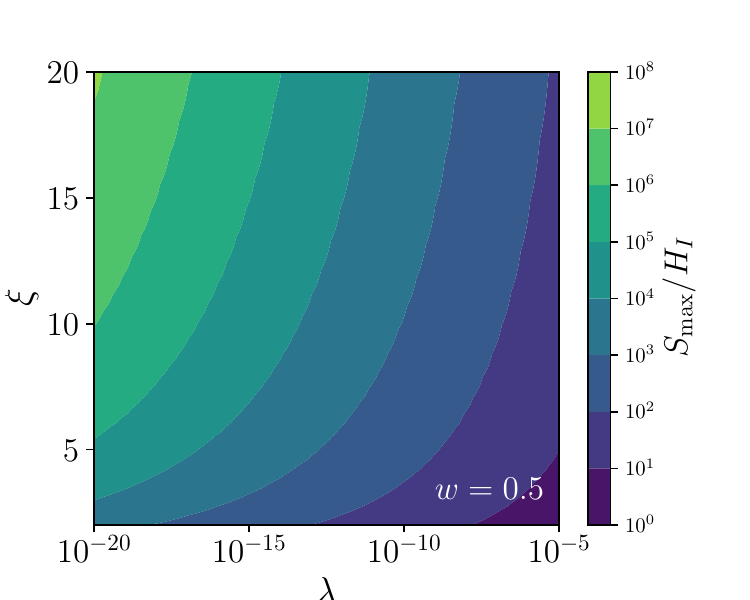}
    \caption{Contour plots of $S_{\text{max}}/H_I$ as function of $\xi,\lambda$ as written in \eqref{Max values complete}, for two values of $w$.}
    \label{fig:Smax}
\end{figure}

\paragraph{Stopping condition and maximum displacement reached}
The maximum value reached by the saxion can be computed as follows. During the tachyonic growth, the saxion will evolve according to Eq.~(\ref{growing solution appendix}) until it reaches the minimum of the potential, which decreases in time according to Eq.~(\ref{Smin}). After this point, the growing will continue until a value $S_\mathrm{max}$, which will be displaced from the minimum by a factor $\mathcal{O}(1)$ at most. Thus, one can estimate $S_\mathrm{max}$ and the Hubble parameter $H_\mathrm{max}$ when this value is reached by equating Eqs.~(\ref{growing solution appendix}) and (\ref{Smin}), obtaining
\begin{subequations}
    \begin{align}
        H_{\text{max}}&\simeq H_I\left(\frac{\lambda}{F_\xi}\right)^{\frac{1}{2+2\alpha}},\\
        S_{\text{max}}&\simeq\sqrt{\frac{3(3w-1)\xi}{\lambda}} H_{\text{max}}=bS_I\left(\frac{F_{\xi}}{\lambda}\right)^{\frac{\alpha}{2+2\alpha}}\,,
        \\
        F_{\xi}&=\frac{36\sqrt{3}\pi^2(3w-1)\xi^{3/2}}{b^2} \,.
        \label{Max values complete}
    \end{align}
\end{subequations}
Equivalently, one can impose that the quartic in the potential and the non-minimal coupling term are of the same order, $-\xi R S^2 = \lambda S^4$, obtaining the same result.
A plot of $S_{\text{max}}/H_I$ for different values of $\xi$ and $\lambda$ is shown in Fig.~\ref{fig:Smax}.
Figure~\ref{fig:FxiGxi}  shows a plot of the functions $F_\xi$ and $G_\xi$, defined in Eq.~\eqref{eq:Gxi} below.
\begin{figure}
    \centering
    \includegraphics[width=0.7\linewidth]{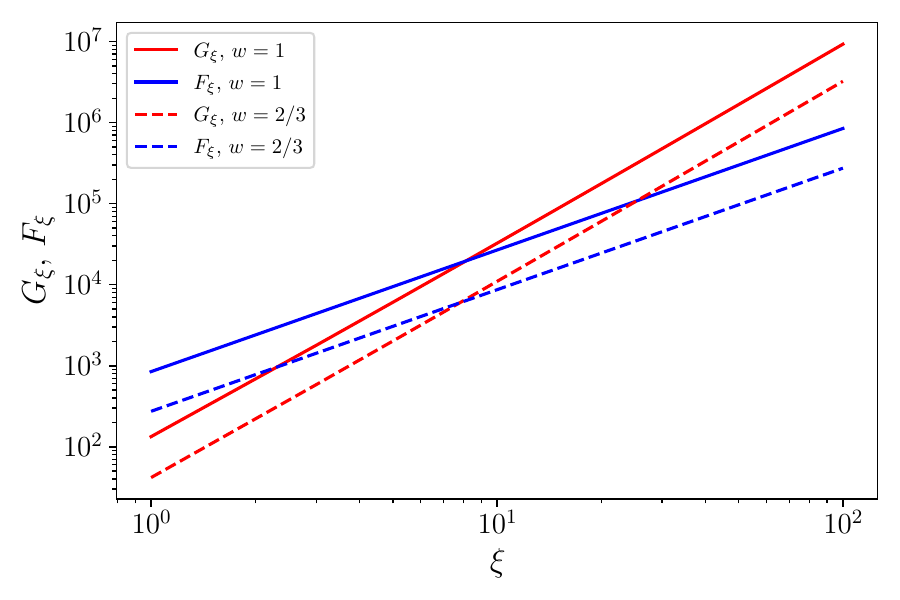}
    \caption{Plot of $F_\xi$ and $G_\xi$, defined in Eq.~\eqref{Max values complete} and \eqref{eq:Gxi}, respectively. Shown is the value for $w=1$ and $w=2/3$. For $w=1/3$ both functions are identically zero.}
    \label{fig:FxiGxi}
\end{figure}

The energy density in the scalar field, neglecting gradients and kinetic energy, is dominated by the term coming from non-minimal coupling for
\begin{equation}
    \frac{H}{H_I} > \left(\frac{b^2\lambda}{6\pi^2(12\xi)^{3/2}}\right)^{\frac{1}{2+2\alpha}}
\end{equation}
which, in particular, is always satisfied at $H=H_\mathrm{max}$.

\paragraph{Gravitational back-reaction}
The tachionic growth of the saxion is only possible as long as the energy is dominated by the inflaton.
We impose the condition that the energy density of the saxion is subdominant in this phase, $\rho_{S,\mathrm{max}} < \frac32 M^2_\mathrm{Pl}H^2_\mathrm{max}$, with $\rho_{S}$ given in Eq.~\ref{saxion energy density}.
We obtain
\begin{align}
    H_{\text{max}}^2 & < \frac{3}{2}\frac{\lambda}{G_\xi}M_\mathrm{Pl}^2\\
    S_{\text{max}}^2 & < \frac{9(3w-1)\xi}{2G_\xi}M_\mathrm{Pl}^2 \quad \overset{w=1}{=} \quad \frac{M_\mathrm{Pl}^2}{(5+4\sqrt{6\xi})\xi} \\
    \lambda & > \left(\frac{2G_\xi H_I^2}{3M_\mathrm{Pl}^2}\right)^\frac{1+\alpha}{\alpha}\frac{1}{F_\xi^{1/\alpha}} \label{gravitaion back reaction generic}
\end{align}
where we defined $G_\xi$ such that
\begin{equation}
    \rho_\mathrm{max} = \frac{G_\xi}{\lambda}H_\mathrm{max}^4 \,,
\end{equation}
hence
\begin{align}
    G_\xi & = \left[\frac{\alpha^2(1+w)}{8\xi}+\alpha+\frac{1}{4}\right] 27(1+w)(3w-1) \xi \label{eq:Gxi} \\
    & = 9(5+4\sqrt{6\xi})\xi^2 \qquad \text{for}\quad w = 1
\end{align}
\paragraph{$\Phi$ field dominance and end of kination}
In Section \ref{sec:Initial Field Dynamics} we computed the duration of kination in the scenario in which the saxion dominates the energy density. Here we write down explicitly the same expressions, showing the dependence on $w$
\begin{equation}
    e^{\Delta N_\mathrm{ke}} = \left(\frac{4\lambda}{3(3w-1)^2\xi^2} \frac{M_\mathrm{Pl}^2}{H_\mathrm{max}^2}\right)^\frac{1}{3w-1}
    \stackrel{w=1}{=} \frac{M_\mathrm{Pl}}{H_\mathrm{max}} \frac{\lambda^{1/2}}{\sqrt{3}\,\xi}\left(2^53^{5/2}(\sqrt{2}-1)^2\pi^2\frac{\xi^{3/2}}{\lambda}\right)^{\frac{3}{2(3+\sqrt{6\xi})}}
\end{equation}
and
\begin{align}
    H_\mathrm{ke} & = H_\mathrm{max} \left(\frac{2\lambda^{1/2}}{\sqrt{3}(3w-1)\xi} \frac{M_\mathrm{Pl}}{H_\mathrm{max}}\right)^{-\frac{3(1+w)}{3w-1}} \\
    & \stackrel{w=1}{=} \frac{3\sqrt{3}\xi^3 H_\mathrm{max}^4}{\lambda^{3/2}M_\mathrm{Pl}^3} 
\end{align}
where, in both equations, in the second step we imposed $w=1$.

\section{Asymmetry factor}\label{asymmetry factor}
In this appendix, we want to give more details about the relation between $\epsilon$ and the orbit shape in field space.
A known result in classical mechanics states that bounded orbits in our potential are not, in general, closed~\cite{2002clme.book.....G}. The quartic potential still admits an exact solution in terms of elliptical functions, and closed orbits do exist~\cite{1980Ap&SS..72...33B}. To get an idea of the relevance of the parameter $\epsilon$, especially in the vicinity of $\epsilon\approx 1$, we find it more instructive to neglect these complications and compare it to the eccentricity of an elliptic orbit, defined as
\begin{equation}
e=\frac{S_{\text{maj}}-S_{\text{min}}}{S_{\text{maj}}+S_{\text{min}}},
\label{eccentricity}
\end{equation}
where $S_{\text{maj}}$ and $S_{\text{min}}$ are, respectively, the value of semi-major and semi-minor axes. This quantity is bounded within $0<e<1$, as $e=0$ reproduces a perfect circular motion and $e=1$ a purely radial oscillation.

Taking a polynomial potential
\begin{equation}
    V=\frac{\lambda}{p}S^p,
\end{equation}
the asymmetry factor is given by
\begin{equation}
    \epsilon=
    \frac{n_{\theta}}{S^2 \sqrt{V'/S}}=\frac{\dot\theta_{\text{maj}}}{\sqrt{\lambda\,S_{\text{maj}}^{p-2}}}
\end{equation}
where all quantities have to be computed at the apoapsis of the orbit, which means $S_{\text{max}}=S_{\text{maj}}$.
Equivalently, $\epsilon$ can be rewritten as the ratio $n_\theta/(p\,n_S)$, with $n_S= \rho_S / m_{S,V'}$, $m_{S,V'} = \sqrt{V'/S}$ and $p$ is the exponent of a generic potential $V\sim S^p$. The term $m_{S,V'}$ enters the equation of motion for the zero-mode of $S$, rather than the usual mass $V''$~\cite{Eroncel:2024rpe}. The advantage of this rewriting is to make explicit the relation of $\epsilon$ to the excitations in the angular/radial directions.

In flat space, because of energy and charge conservation, we can write down the following relations
\begin{align}
    \frac{1}{2}S_{\text{maj}}^2\dot\theta_{\text{maj}}^2+\frac{\lambda}{p}S_{\text{maj}}^p&=\frac{1}{2}S_{\text{min}}^2\dot\theta_{\text{min}}^2+\frac{\lambda}{p}S_{\text{min}}^p,\\
    S^2_{\text{maj}}\dot\theta_{\text{maj}}&=S^2_{\text{min}}\dot\theta_{\text{min}},
\end{align}
combining this two equations and using the definition \eqref{epsilon def} we obtain
\begin{equation}
    \epsilon^2=\frac{2}{p}\frac{S_{\text{min}}^2}{S_{\text{maj}}^p}\left(\frac{S_{\text{maj}}^p-S_{\text{min}}^p}{S_{\text{maj}}^2-S_{\text{min}}^2}\right),
    \label{epsilon per S}
\end{equation}
comparing \eqref{epsilon per S} with \eqref{eccentricity} we get the following relation
\begin{equation}
    \epsilon^2=\frac{(e-1)^2}{2pe}\left(1-\left(\frac{1-e}{1+e}\right)^p\right)\,,
\end{equation}
wich in the limit $e\ll 1$ becomes
\begin{equation}
    \epsilon^2\simeq 1 - (2+p)e + \mathcal{O}(e^2),
\end{equation}
this means that for a perfectly circular orbit $e\rightarrow0$, $\epsilon\rightarrow1$. Conversely, expanding for $1-e\approx0$ we obtain $\epsilon\to 0$ as expected, although in this limit the deviation from a closed, elliptical orbit is large, and the comparison made above looses its significance.

For a nearly circular orbit, and using the conservation of angular momentum, it is easy to show that $\dot\theta_\mathrm{maj} = \dot\theta_\mathrm{max} [1+\mathcal{O}(e)]$, and it is thus safe to use $\dot\theta_\mathrm{max}$ when computing $\epsilon$.

\section{Axion kick}\label{Axion kick}
\begin{figure}[htb]
    \centering
    \includegraphics[width=0.8\linewidth]{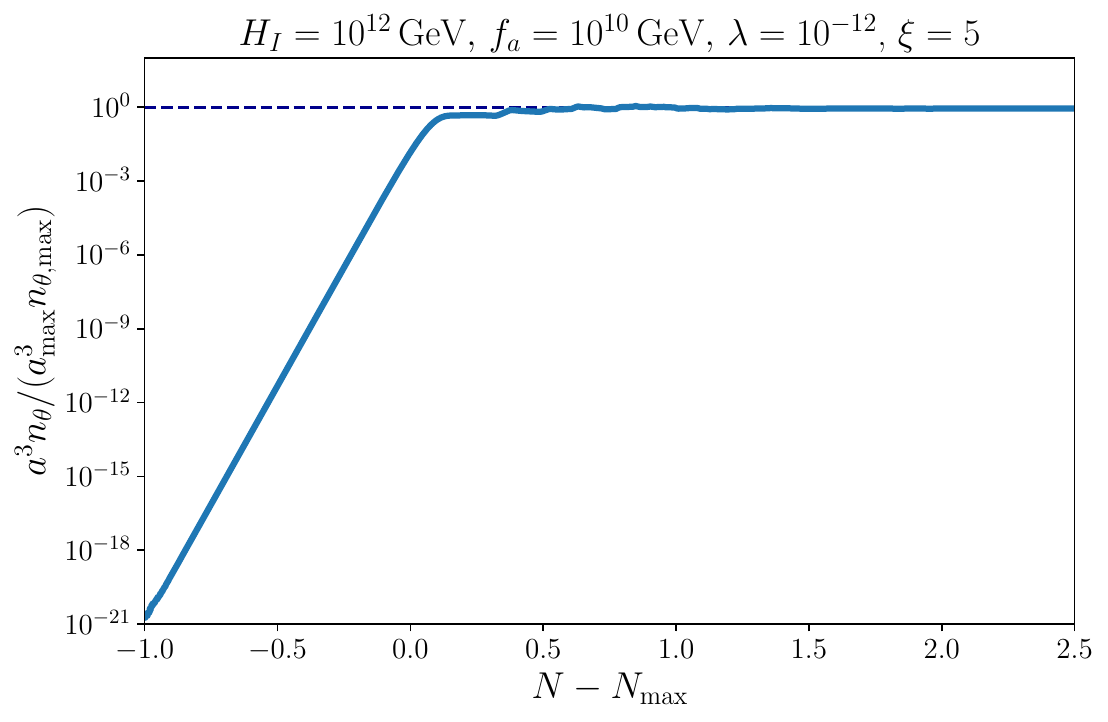}
    \caption{The blue line shows the evolution of $n_{\theta}$ in a comoving volume during the tachyonic growth and after the kick is imparted, the dashed line is at 1, showing a matching with the analytic result of Eq.~\eqref{ntheta integrated}.}
    \label{nTheta}
\end{figure}

We can explicitly compute the produced $n_{\theta}$ by integrating Eq.~\eqref{PQ charge}, we rewrite it here
\begin{equation}
    \frac{\text{d}}{\text{d}t}\left(a^3n_{\theta}\right)=a^32A\frac{S^n}{M_{\text{Pl}}^{n-3}}\sin(n\theta)
\end{equation}
integrating over $a$ we get 
\begin{subequations}
\begin{align}
    a^3n_{\theta}-a^3_{I}n_{\theta,I}&=\int_{a_I}^{a}\text{d}a'\,\frac{a'^2}{H}\frac{2AS^n}{M_{\text{Pl}}^{n-3}}\sin(n\theta)\\
    &\simeq 2A\frac{S_{\text{max}}^n}{M_{\text{Pl}}^{n-3}}\sin(n\theta_I)\left[H_{\text{max}}^{n\alpha}\int_{a_I}^{a_{\text{max}}}\text{d}a'\,\frac{a'^2}{H^{n\alpha+1}}+a_{\text{max}}^{\frac{6n}{p+2}}\int_{a_{\text{max}}}^{a}\text{d}a'\frac{a'^{\frac{2(p+2)-6n}{p+2}}}{H}\right]
    \label{integral charge}
\end{align}
\end{subequations}
where in the first integral we plugged in the growing solution obtained in Eq.~\eqref{growing solution} 
\begin{equation}
    S(t)=S_{\text{max}}H_{\text{max}}^{\alpha}/H(t)^{\alpha},
\end{equation}
and in the second one
\begin{equation}
    S(t)=S_{\text{max}}\left(\frac{a_{\text{max}}}{a}\right)^{\frac{6}{p+2}},
\end{equation}
which is the typical redshift of an oscillating field with a potential $V\sim \lambda S^p$~\cite{Turner:1983he}.
If $4n>(2+p)(3+w)$, the second integral is convergent in the limit $a\rightarrow\infty$. This reflects the conservation of the charge at late times, once the $U(1)$ breaking potential is no longer relevant. After taking the limits $a\gg a_\mathrm{max}$, $H_I/H_{\text{max}}\gg1$ and assuming $n_{\theta,I}\simeq0$, the integral of Eq.~\eqref{integral charge} gives
\begin{equation}
    a_\infty^3 n_{\theta,\infty}\simeq \beta\sin{(n\theta_I)}a_I^3\left(\frac{H_I}{H_{\text{max}}}\right)^{\frac{3+w}{1+w}}\frac{A S_{\text{max}}^n}{M_{\text{Pl}}^{n-3}H_I},
\end{equation}
The factor $\beta$ is given by
\begin{equation}
    \beta=\frac{4n[4+\alpha(p+2)(w+1))}{3[w+3+n\alpha(w+1)][4n-(p+2)(w+3)]}
    \approx \frac{4(p+2)}{12n-3(p+2)(3+w)} \,,
\end{equation}
where, in the last step, we took the limit $\xi\gg 1$.
For simplicity, one can pretend this charge is immediately generated at the kick and is further conserved up to expansion, so by redshifting this quantity back to $a_{\text{max}}$ we get
\begin{equation}
    n_{\theta,\text{max}}\simeq\beta \sin(n\theta_I)\frac{AS_{\text{max}}^n}{M_{\text{Pl}}^{n-3}H_{\text{max}}}
    \label{ntheta integrated}
\end{equation}
Although this value does not reproduce the charge at the time $a_\mathrm{max}$, when redshifted to late times the correct amount of charge is obtained: $n_\theta = n_{\theta,\text{max}} (a_\mathrm{max}/a)^3$.
A plot of the evolution of the comoving charge computed numerically throughout the kick for $w=1$, $p=4$, $n=7$ is shown in Fig.~\ref{nTheta}.

The axion field velocity is obtained from $n_\theta$:
\begin{equation}
   \dot\theta_{\text{max}}=\beta \sin(n\theta_I)\frac{AS_{\text{max}}^{n-2}}{M_{\text{Pl}}^{n-3}H_{\text{max}}}.
\end{equation}
and from the definition of $\epsilon$ introduce in Eq.~\eqref{epsilon def}, we obtain
\begin{equation}\label{eq:epsilon}
    \epsilon\simeq \beta\sin(n\theta_I)\frac{A}{\sqrt{\lambda} H_{\text{max}}}\left(\frac{S_{\text{max}}}{M_{\text{Pl}}}\right)^{n-3}\,,
\end{equation}
where we identified $S_\mathrm{maj}\simeq S_\mathrm{max}$ and $\dot\theta_\mathrm{maj}\simeq\dot\theta_\mathrm{max}$.
Kinetic misalignment is efficient only if $\epsilon\sim\mathcal{O}(1)$, in order to avoid parametric resonance effects. This condition does not come out naturally, in particular we need to set the coupling $A$ in order to respect Eq.~\ref{eq:epsilon}.
In practice, we may ask  the averaged value on different domains to be $|\bar{\epsilon}|=0.5$ and fix
\begin{equation}\label{A}
    A\simeq \sqrt\frac{{\lambda}}{2\beta^2} H_{\text{max}}\left(\frac{M_{\text{Pl}}}{S_{\text{max}}}\right)^{n-3}.
\end{equation}
Equation~\ref{A} should be understood as a coincidence of scales, rather than an exact tuning. If $A$ is smaller than this value, the kick will be insufficient for a circular orbit. If $A$ is instead larger, the field will be pushed to even larger value of $S$, again resulting in an elongated orbit. In both cases, we expect $\epsilon\lesssim 1$ as long as $A$ is close to the value in Eq.~\ref{A}. Obtaining $\epsilon = 1$ exactly is not possible in the presence of the $U(1)$ violating term in the potential.
Requiring $A<M_\mathrm{Pl}$ we obtain another upper bound on the quartic coupling $\lambda$:
\begin{equation}
    \lambda<\left[\sqrt{2}\beta (3(3\omega-1)\xi)^{\frac{n-3}{2}}\left(\frac{H_I}{M_{\text{Pl}}}\right)^{n-4}F^{\frac{4-n}{2+2\alpha}}\right]^{\frac{2+2\alpha}{2+(n-2)\alpha}}.
    \label{boundA}
\end{equation}

\clearpage

\bibliographystyle{JHEP}
\bibliography{biblio}

\providecommand{\href}[2]{#2}\begingroup\raggedright\begin{thebibliography}{10}

\bibitem{Preskill:1982cy}
J.~Preskill, M.B.~Wise and F.~Wilczek, \emph{{Cosmology of the Invisible
  Axion}}, \href{https://doi.org/10.1016/0370-2693(83)90637-8}{\emph{Phys.
  Lett. B} {\bfseries 120} (1983) 127}.

\bibitem{Abbott:1982af}
L.F.~Abbott and P.~Sikivie, \emph{{A Cosmological Bound on the Invisible
  Axion}}, \href{https://doi.org/10.1016/0370-2693(83)90638-X}{\emph{Phys.
  Lett. B} {\bfseries 120} (1983) 133}.

\bibitem{Dine:1982ah}
M.~Dine and W.~Fischler, \emph{{The Not So Harmless Axion}},
  \href{https://doi.org/10.1016/0370-2693(83)90639-1}{\emph{Phys. Lett. B}
  {\bfseries 120} (1983) 137}.

\bibitem{Gorghetto:2018myk}
M.~Gorghetto, E.~Hardy and G.~Villadoro, \emph{{Axions from Strings: the
  Attractive Solution}},
  \href{https://doi.org/10.1007/JHEP07(2018)151}{\emph{JHEP} {\bfseries 07}
  (2018) 151} [\href{https://arxiv.org/abs/1806.04677}{{\ttfamily
  1806.04677}}].

\bibitem{Gorghetto:2020qws}
M.~Gorghetto, E.~Hardy and G.~Villadoro, \emph{{More axions from strings}},
  \href{https://doi.org/10.21468/SciPostPhys.10.2.050}{\emph{SciPost Phys.}
  {\bfseries 10} (2021) 050}
  [\href{https://arxiv.org/abs/2007.04990}{{\ttfamily 2007.04990}}].

\bibitem{Gorghetto:2021fsn}
M.~Gorghetto, E.~Hardy and H.~Nicolaescu, \emph{{Observing invisible axions
  with gravitational waves}},
  \href{https://doi.org/10.1088/1475-7516/2021/06/034}{\emph{JCAP} {\bfseries
  06} (2021) 034} [\href{https://arxiv.org/abs/2101.11007}{{\ttfamily
  2101.11007}}].

\bibitem{Baumholzer:2020hvx}
S.~Baumholzer, V.~Brdar and E.~Morgante, \emph{{Structure Formation Limits on
  Axion-Like Dark Matter}},
  \href{https://doi.org/10.1088/1475-7516/2021/05/004}{\emph{JCAP} {\bfseries
  05} (2021) 004} [\href{https://arxiv.org/abs/2012.09181}{{\ttfamily
  2012.09181}}].

\bibitem{Becker:2025yvb}
M.~Becker, J.~Harz, E.~Morgante, C.~Puchades-Ib{\'a}{\~n}ez and P.~Schwaller,
  \emph{{ALP production from abelian gauge bosons: beyond hard thermal loops}},
  \href{https://doi.org/10.1007/JHEP06(2025)160}{\emph{JHEP} {\bfseries 06}
  (2025) 160} [\href{https://arxiv.org/abs/2502.01729}{{\ttfamily
  2502.01729}}].

\bibitem{Co:2019jts}
R.T.~Co, L.J.~Hall and K.~Harigaya, \emph{{Axion Kinetic Misalignment
  Mechanism}},
  \href{https://doi.org/10.1103/PhysRevLett.124.251802}{\emph{Phys. Rev. Lett.}
  {\bfseries 124} (2020) 251802}
  [\href{https://arxiv.org/abs/1910.14152}{{\ttfamily 1910.14152}}].

\bibitem{Chang:2019tvx}
C.-F.~Chang and Y.~Cui, \emph{{New Perspectives on Axion Misalignment
  Mechanism}}, \href{https://doi.org/10.1103/PhysRevD.102.015003}{\emph{Phys.
  Rev. D} {\bfseries 102} (2020) 015003}
  [\href{https://arxiv.org/abs/1911.11885}{{\ttfamily 1911.11885}}].

\bibitem{Co:2020dya}
R.T.~Co, L.J.~Hall, K.~Harigaya, K.A.~Olive and S.~Verner, \emph{{Axion Kinetic
  Misalignment and Parametric Resonance from Inflation}},
  \href{https://doi.org/10.1088/1475-7516/2020/08/036}{\emph{JCAP} {\bfseries
  08} (2020) 036} [\href{https://arxiv.org/abs/2004.00629}{{\ttfamily
  2004.00629}}].

\bibitem{Kobayashi:2020bxq}
T.~Kobayashi and R.K.~Jain, \emph{{Impact of Helical Electromagnetic Fields on
  the Axion Window}},
  \href{https://doi.org/10.1088/1475-7516/2021/03/025}{\emph{JCAP} {\bfseries
  03} (2021) 025} [\href{https://arxiv.org/abs/2012.00896}{{\ttfamily
  2012.00896}}].

\bibitem{Domcke:2022wpb}
V.~Domcke, K.~Harigaya and K.~Mukaida, \emph{{Charge transfer between rotating
  complex scalar fields}},
  \href{https://doi.org/10.1007/JHEP08(2022)234}{\emph{JHEP} {\bfseries 08}
  (2022) 234} [\href{https://arxiv.org/abs/2205.00942}{{\ttfamily
  2205.00942}}].

\bibitem{Co:2023mhe}
R.T.~Co and M.~Yamada, \emph{{Axion cogenesis without isocurvature
  perturbations}},
  \href{https://doi.org/10.1103/PhysRevD.110.055009}{\emph{Phys. Rev. D}
  {\bfseries 110} (2024) 055009}
  [\href{https://arxiv.org/abs/2312.17730}{{\ttfamily 2312.17730}}].

\bibitem{Lee:2023dtw}
H.M.~Lee, A.G.~Menkara, M.-J.~Seong and J.-H.~Song, \emph{{Peccei-Quinn
  inflation at the pole and axion kinetic misalignment}},
  \href{https://doi.org/10.1007/JHEP05(2024)295}{\emph{JHEP} {\bfseries 05}
  (2024) 295} [\href{https://arxiv.org/abs/2310.17710}{{\ttfamily
  2310.17710}}].

\bibitem{Lee:2024bij}
H.M.~Lee, A.G.~Menkara, M.-J.~Seong and J.-H.~Song, \emph{{Inflation models
  with Peccei{\textendash}Quinn symmetry and axion kinetic misalignment}},
  \href{https://doi.org/10.1140/epjc/s10052-024-13648-y}{\emph{Eur. Phys. J. C}
  {\bfseries 84} (2024) 1260}
  [\href{https://arxiv.org/abs/2408.17013}{{\ttfamily 2408.17013}}].

\bibitem{Eroncel:2024rpe}
C.~Er{\"o}ncel, R.~Sato, G.~Servant and P.~S{\o}rensen, \emph{{Model
  implementations of axion dark matter from kinetic misalignment}},
  \href{https://doi.org/10.1088/1475-7516/2025/08/087}{\emph{JCAP} {\bfseries
  08} (2025) 087} [\href{https://arxiv.org/abs/2408.08355}{{\ttfamily
  2408.08355}}].

\bibitem{Co:2019wyp}
R.T.~Co and K.~Harigaya, \emph{{Axiogenesis}},
  \href{https://doi.org/10.1103/PhysRevLett.124.111602}{\emph{Phys. Rev. Lett.}
  {\bfseries 124} (2020) 111602}
  [\href{https://arxiv.org/abs/1910.02080}{{\ttfamily 1910.02080}}].

\bibitem{Co:2020jtv}
R.T.~Co, N.~Fernandez, A.~Ghalsasi, L.J.~Hall and K.~Harigaya,
  \emph{{Lepto-Axiogenesis}},
  \href{https://doi.org/10.1007/JHEP03(2021)017}{\emph{JHEP} {\bfseries 03}
  (2021) 017} [\href{https://arxiv.org/abs/2006.05687}{{\ttfamily
  2006.05687}}].

\bibitem{Madge:2021abk}
E.~Madge, W.~Ratzinger, D.~Schmitt and P.~Schwaller, \emph{{Audible axions with
  a booster: Stochastic gravitational waves from rotating ALPs}},
  \href{https://doi.org/10.21468/SciPostPhys.12.5.171}{\emph{SciPost Phys.}
  {\bfseries 12} (2022) 171}
  [\href{https://arxiv.org/abs/2111.12730}{{\ttfamily 2111.12730}}].

\bibitem{Co:2021rhi}
R.T.~Co, K.~Harigaya and A.~Pierce, \emph{{Gravitational waves and dark photon
  dark matter from axion rotations}},
  \href{https://doi.org/10.1007/JHEP12(2021)099}{\emph{JHEP} {\bfseries 12}
  (2021) 099} [\href{https://arxiv.org/abs/2104.02077}{{\ttfamily
  2104.02077}}].

\bibitem{Harigaya:2021txz}
K.~Harigaya and I.R.~Wang, \emph{{Axiogenesis from $SU(2)_R$ phase
  transition}}, \href{https://doi.org/10.1007/JHEP10(2021)022}{\emph{JHEP}
  {\bfseries 10} (2021) 022}
  [\href{https://arxiv.org/abs/2107.09679}{{\ttfamily 2107.09679}}].

\bibitem{Co:2021lkc}
R.T.~Co, D.~Dunsky, N.~Fernandez, A.~Ghalsasi, L.J.~Hall, K.~Harigaya et~al.,
  \emph{{Gravitational wave and CMB probes of axion kination}},
  \href{https://doi.org/10.1007/JHEP09(2022)116}{\emph{JHEP} {\bfseries 09}
  (2022) 116} [\href{https://arxiv.org/abs/2108.09299}{{\ttfamily
  2108.09299}}].

\bibitem{Co:2021qgl}
R.T.~Co, K.~Harigaya, Z.~Johnson and A.~Pierce, \emph{{R-parity violation
  axiogenesis}}, \href{https://doi.org/10.1007/JHEP11(2021)210}{\emph{JHEP}
  {\bfseries 11} (2021) 210}
  [\href{https://arxiv.org/abs/2110.05487}{{\ttfamily 2110.05487}}].

\bibitem{Gouttenoire:2021wzu}
Y.~Gouttenoire, G.~Servant and P.~Simakachorn, \emph{{Revealing the Primordial
  Irreducible Inflationary Gravitational-Wave Background with a Spinning
  Peccei-Quinn Axion}},  \href{https://arxiv.org/abs/2108.10328}{{\ttfamily
  2108.10328}}.

\bibitem{Gouttenoire:2021jhk}
Y.~Gouttenoire, G.~Servant and P.~Simakachorn, \emph{{Kination cosmology from
  scalar fields and gravitational-wave signatures}},
  \href{https://arxiv.org/abs/2111.01150}{{\ttfamily 2111.01150}}.

\bibitem{Barnes:2022ren}
P.~Barnes, R.T.~Co, K.~Harigaya and A.~Pierce, \emph{{Lepto-axiogenesis and the
  scale of supersymmetry}},
  \href{https://doi.org/10.1007/JHEP05(2023)114}{\emph{JHEP} {\bfseries 05}
  (2023) 114} [\href{https://arxiv.org/abs/2208.07878}{{\ttfamily
  2208.07878}}].

\bibitem{Co:2022qpr}
R.T.~Co, K.~Harigaya and A.~Pierce, \emph{{Cosmic perturbations from a rotating
  field}}, \href{https://doi.org/10.1088/1475-7516/2022/10/037}{\emph{JCAP}
  {\bfseries 10} (2022) 037}
  [\href{https://arxiv.org/abs/2202.01785}{{\ttfamily 2202.01785}}].

\bibitem{Co:2022aav}
R.T.~Co, T.~Gherghetta and K.~Harigaya, \emph{{Axiogenesis with a heavy QCD
  axion}}, \href{https://doi.org/10.1007/JHEP10(2022)121}{\emph{JHEP}
  {\bfseries 10} (2022) 121}
  [\href{https://arxiv.org/abs/2206.00678}{{\ttfamily 2206.00678}}].

\bibitem{Barnes:2024jap}
P.~Barnes, R.T.~Co, K.~Harigaya and A.~Pierce, \emph{{Lepto-axiogenesis with
  light right-handed neutrinos}},
  \href{https://doi.org/10.1007/JHEP08(2025)004}{\emph{JHEP} {\bfseries 08}
  (2025) 004} [\href{https://arxiv.org/abs/2402.10263}{{\ttfamily
  2402.10263}}].

\bibitem{Bodas:2025wef}
A.~Bodas, K.~Harigaya, K.~Inomata, T.~Terada and L.-T.~Wang, \emph{{Anisotropic
  Gravitational Waves from Anisotropic Axion Rotation}},
  \href{https://arxiv.org/abs/2508.08249}{{\ttfamily 2508.08249}}.

\bibitem{Co:2025lrd}
R.T.~Co, N.~Fernandez, A.~Ghalsasi, K.~Harigaya and J.~Shelton, \emph{{Enhanced
  Matter Power Spectrum from Axion Kination after Big Bang Nucleosynthesis}},
  \href{https://arxiv.org/abs/2510.01308}{{\ttfamily 2510.01308}}.

\bibitem{Eroncel:2025qlk}
C.~Er{\"o}ncel, Y.~Gouttenoire, R.~Sato, G.~Servant and P.~Simakachorn,
  \emph{{New Source for QCD Axion Dark Matter Production: Curvature Induced}},
  \href{https://doi.org/10.1103/s3t7-41t2}{\emph{Phys. Rev. Lett.} {\bfseries
  135} (2025) 231002} [\href{https://arxiv.org/abs/2503.04880}{{\ttfamily
  2503.04880}}].

\bibitem{Bodas:2025eca}
A.~Bodas, R.T.~Co, A.~Ghalsasi, K.~Harigaya and L.-T.~Wang, \emph{{Acoustic
  misalignment mechanism for axion dark matter}},
  \href{https://doi.org/10.1007/JHEP08(2025)131}{\emph{JHEP} {\bfseries 08}
  (2025) 131} [\href{https://arxiv.org/abs/2503.04888}{{\ttfamily
  2503.04888}}].

\bibitem{Spokoiny:1993kt}
B.~Spokoiny, \emph{{Deflationary universe scenario}},
  \href{https://doi.org/10.1016/0370-2693(93)90155-B}{\emph{Phys. Lett. B}
  {\bfseries 315} (1993) 40}
  [\href{https://arxiv.org/abs/gr-qc/9306008}{{\ttfamily gr-qc/9306008}}].

\bibitem{Peebles:1998qn}
P.J.E.~Peebles and A.~Vilenkin, \emph{{Quintessential inflation}},
  \href{https://doi.org/10.1103/PhysRevD.59.063505}{\emph{Phys. Rev. D}
  {\bfseries 59} (1999) 063505}
  [\href{https://arxiv.org/abs/astro-ph/9810509}{{\ttfamily
  astro-ph/9810509}}].

\bibitem{Akrami:2017cir}
Y.~Akrami, R.~Kallosh, A.~Linde and V.~Vardanyan, \emph{{Dark energy,
  $\alpha$-attractors, and large-scale structure surveys}},
  \href{https://doi.org/10.1088/1475-7516/2018/06/041}{\emph{JCAP} {\bfseries
  06} (2018) 041} [\href{https://arxiv.org/abs/1712.09693}{{\ttfamily
  1712.09693}}].

\bibitem{Dimopoulos:2019ogl}
K.~Dimopoulos, M.~Kar{\v{c}}iauskas and C.~Owen, \emph{{Quintessential
  inflation with a trap and axionic dark matter}},
  \href{https://doi.org/10.1103/PhysRevD.100.083530}{\emph{Phys. Rev. D}
  {\bfseries 100} (2019) 083530}
  [\href{https://arxiv.org/abs/1907.04676}{{\ttfamily 1907.04676}}].

\bibitem{Bettoni:2018utf}
D.~Bettoni and J.~Rubio, \emph{{Quintessential Affleck-Dine baryogenesis with
  non-minimal couplings}},
  \href{https://doi.org/10.1016/j.physletb.2018.07.046}{\emph{Phys. Lett. B}
  {\bfseries 784} (2018) 122}
  [\href{https://arxiv.org/abs/1805.02669}{{\ttfamily 1805.02669}}].

\bibitem{Bettoni:2019dcw}
D.~Bettoni and J.~Rubio, \emph{{Hubble-induced phase transitions: Walls are not
  forever}}, \href{https://doi.org/10.1088/1475-7516/2020/01/002}{\emph{JCAP}
  {\bfseries 01} (2020) 002}
  [\href{https://arxiv.org/abs/1911.03484}{{\ttfamily 1911.03484}}].

\bibitem{Bettoni:2021zhq}
D.~Bettoni, A.~Lopez-Eiguren and J.~Rubio, \emph{{Hubble-induced phase
  transitions on the lattice with applications to Ricci reheating}},
  \href{https://doi.org/10.1088/1475-7516/2022/01/002}{\emph{JCAP} {\bfseries
  01} (2022) 002} [\href{https://arxiv.org/abs/2107.09671}{{\ttfamily
  2107.09671}}].

\bibitem{Bettoni:2024ixe}
D.~Bettoni, G.~Laverda, A.~Lopez-Eiguren and J.~Rubio, \emph{{Hubble-induced
  phase transitions: gravitational-wave imprint of Ricci reheating from lattice
  simulations}},
  \href{https://doi.org/10.1088/1475-7516/2025/03/027}{\emph{JCAP} {\bfseries
  03} (2025) 027} [\href{https://arxiv.org/abs/2409.15450}{{\ttfamily
  2409.15450}}].

\bibitem{Opferkuch:2019zbd}
T.~Opferkuch, P.~Schwaller and B.A.~Stefanek, \emph{{Ricci Reheating}},
  \href{https://doi.org/10.1088/1475-7516/2019/07/016}{\emph{JCAP} {\bfseries
  07} (2019) 016} [\href{https://arxiv.org/abs/1905.06823}{{\ttfamily
  1905.06823}}].

\bibitem{Laverda:2023uqv}
G.~Laverda and J.~Rubio, \emph{{Ricci reheating reloaded}},
  \href{https://doi.org/10.1088/1475-7516/2024/03/033}{\emph{JCAP} {\bfseries
  03} (2024) 033} [\href{https://arxiv.org/abs/2307.03774}{{\ttfamily
  2307.03774}}].

\bibitem{Figueroa:2024asq}
D.G.~Figueroa, T.~Opferkuch and B.A.~Stefanek, \emph{{Ricci Reheating on the
  Lattice}},  \href{https://arxiv.org/abs/2404.17654}{{\ttfamily 2404.17654}}.

\bibitem{Fedderke:2025sic}
M.A.~Fedderke, J.~Huang and N.~Siemonsen, \emph{{Periodic cosmic string
  formation and dynamics}},
  \href{https://doi.org/10.1007/JHEP08(2025)080}{\emph{JHEP} {\bfseries 08}
  (2025) 080} [\href{https://arxiv.org/abs/2503.03116}{{\ttfamily
  2503.03116}}].

\bibitem{Fonseca:2019ypl}
N.~Fonseca, E.~Morgante, R.~Sato and G.~Servant, \emph{{Axion fragmentation}},
  \href{https://doi.org/10.1007/JHEP04(2020)010}{\emph{JHEP} {\bfseries 04}
  (2020) 010} [\href{https://arxiv.org/abs/1911.08472}{{\ttfamily
  1911.08472}}].

\bibitem{Morgante:2021bks}
E.~Morgante, W.~Ratzinger, R.~Sato and B.A.~Stefanek, \emph{{Axion
  fragmentation on the lattice}},
  \href{https://doi.org/10.1007/JHEP12(2021)037}{\emph{JHEP} {\bfseries 12}
  (2021) 037} [\href{https://arxiv.org/abs/2109.13823}{{\ttfamily
  2109.13823}}].

\bibitem{Eroncel:2022vjg}
C.~Er{\"o}ncel, R.~Sato, G.~Servant and P.~S{\o}rensen, \emph{{ALP dark matter
  from kinetic fragmentation: opening up the parameter window}},
  \href{https://doi.org/10.1088/1475-7516/2022/10/053}{\emph{JCAP} {\bfseries
  10} (2022) 053} [\href{https://arxiv.org/abs/2206.14259}{{\ttfamily
  2206.14259}}].

\bibitem{Eroncel:2022efc}
C.~Er{\"o}ncel and G.~Servant, \emph{{ALP dark matter mini-clusters from
  kinetic fragmentation}},
  \href{https://doi.org/10.1088/1475-7516/2023/01/009}{\emph{JCAP} {\bfseries
  01} (2023) 009} [\href{https://arxiv.org/abs/2207.10111}{{\ttfamily
  2207.10111}}].

\bibitem{Fasiello:2025ptb}
M.~Fasiello, J.~Lizarraga, A.~Papageorgiou and A.~Urio, \emph{{Kinetic
  fragmentation of the QCD axion on the lattice}},
  \href{https://doi.org/10.1088/1475-7516/2025/09/019}{\emph{JCAP} {\bfseries
  09} (2025) 019} [\href{https://arxiv.org/abs/2507.01822}{{\ttfamily
  2507.01822}}].

\bibitem{Kibble:1976sj}
T.W.B.~Kibble, \emph{{Topology of Cosmic Domains and Strings}},
  \href{https://doi.org/10.1088/0305-4470/9/8/029}{\emph{J. Phys. A} {\bfseries
  9} (1976) 1387}.

\bibitem{Vilenkin:1981kz}
A.~Vilenkin, \emph{{Cosmic Strings}},
  \href{https://doi.org/10.1103/PhysRevD.24.2082}{\emph{Phys. Rev. D}
  {\bfseries 24} (1981) 2082}.

\bibitem{Sikivie:1982qv}
P.~Sikivie, \emph{{Of Axions, Domain Walls and the Early Universe}},
  \href{https://doi.org/10.1103/PhysRevLett.48.1156}{\emph{Phys. Rev. Lett.}
  {\bfseries 48} (1982) 1156}.

\bibitem{Vilenkin:1982ks}
A.~Vilenkin and A.E.~Everett, \emph{{Cosmic Strings and Domain Walls in Models
  with Goldstone and PseudoGoldstone Bosons}},
  \href{https://doi.org/10.1103/PhysRevLett.48.1867}{\emph{Phys. Rev. Lett.}
  {\bfseries 48} (1982) 1867}.

\bibitem{Sikivie:2006ni}
P.~Sikivie, \emph{{Axion Cosmology}},
  \href{https://doi.org/10.1007/978-3-540-73518-2_2}{\emph{Lect. Notes Phys.}
  {\bfseries 741} (2008) 19}
  [\href{https://arxiv.org/abs/astro-ph/0610440}{{\ttfamily
  astro-ph/0610440}}].

\bibitem{Eroncel:2025bcb}
C.~Er{\"o}ncel, Y.~Gouttenoire, R.~Sato, G.~Servant and P.~Simakachorn,
  \emph{{Universal Bound on the Duration of a Kination Era}},
  \href{https://doi.org/10.1103/k7ty-gwjg}{\emph{Phys. Rev. Lett.} {\bfseries
  135} (2025) 101002} [\href{https://arxiv.org/abs/2501.17226}{{\ttfamily
  2501.17226}}].

\bibitem{Planck:2018jri}
{\scshape Planck} collaboration, \emph{{Planck 2018 results. X. Constraints on
  inflation}}, \href{https://doi.org/10.1051/0004-6361/201833887}{\emph{Astron.
  Astrophys.} {\bfseries 641} (2020) A10}
  [\href{https://arxiv.org/abs/1807.06211}{{\ttfamily 1807.06211}}].

\bibitem{Correia:2024cpk}
J.~Correia, M.~Hindmarsh, J.~Lizarraga, A.~Lopez-Eiguren, K.~Rummukainen and
  J.~Urrestilla, \emph{{Scaling density of axion strings in terasite
  simulations}}, \href{https://doi.org/10.1103/PhysRevD.111.063532}{\emph{Phys.
  Rev. D} {\bfseries 111} (2025) 063532}
  [\href{https://arxiv.org/abs/2410.18064}{{\ttfamily 2410.18064}}].

\bibitem{Kim:2024wku}
H.~Kim, J.~Park and M.~Son, \emph{{Axion dark matter from cosmic string
  network}}, \href{https://doi.org/10.1007/JHEP07(2024)150}{\emph{JHEP}
  {\bfseries 07} (2024) 150}
  [\href{https://arxiv.org/abs/2402.00741}{{\ttfamily 2402.00741}}].

\bibitem{Kim:2024dtq}
H.~Kim and M.~Son, \emph{{More scalings from cosmic strings}},
  \href{https://doi.org/10.1007/JHEP07(2025)052}{\emph{JHEP} {\bfseries 07}
  (2025) 052} [\href{https://arxiv.org/abs/2411.08455}{{\ttfamily
  2411.08455}}].

\bibitem{Kawasaki:2014sqa}
M.~Kawasaki, K.~Saikawa and T.~Sekiguchi, \emph{{Axion dark matter from
  topological defects}},
  \href{https://doi.org/10.1103/PhysRevD.91.065014}{\emph{Phys. Rev. D}
  {\bfseries 91} (2015) 065014}
  [\href{https://arxiv.org/abs/1412.0789}{{\ttfamily 1412.0789}}].

\bibitem{Saikawa:2024bta}
K.~Saikawa, J.~Redondo, A.~Vaquero and M.~Kaltschmidt, \emph{{Spectrum of
  global string networks and the axion dark matter mass}},
  \href{https://doi.org/10.1088/1475-7516/2024/10/043}{\emph{JCAP} {\bfseries
  10} (2024) 043} [\href{https://arxiv.org/abs/2401.17253}{{\ttfamily
  2401.17253}}].

\bibitem{Buschmann:2021sdq}
M.~Buschmann, J.W.~Foster, A.~Hook, A.~Peterson, D.E.~Willcox, W.~Zhang et~al.,
  \emph{{Dark matter from axion strings with adaptive mesh refinement}},
  \href{https://doi.org/10.1038/s41467-022-28669-y}{\emph{Nature Commun.}
  {\bfseries 13} (2022) 1049}
  [\href{https://arxiv.org/abs/2108.05368}{{\ttfamily 2108.05368}}].

\bibitem{Benabou:2024msj}
J.N.~Benabou, M.~Buschmann, J.W.~Foster and B.R.~Safdi, \emph{{Axion Mass
  Prediction from Adaptive Mesh Refinement Cosmological Lattice Simulations}},
  \href{https://doi.org/10.1103/6v21-d6sj}{\emph{Phys. Rev. Lett.} {\bfseries
  134} (2025) 241003} [\href{https://arxiv.org/abs/2412.08699}{{\ttfamily
  2412.08699}}].

\bibitem{Baker:2006ts}
C.A.~Baker et~al., \emph{{An Improved experimental limit on the electric dipole
  moment of the neutron}},
  \href{https://doi.org/10.1103/PhysRevLett.97.131801}{\emph{Phys. Rev. Lett.}
  {\bfseries 97} (2006) 131801}
  [\href{https://arxiv.org/abs/hep-ex/0602020}{{\ttfamily hep-ex/0602020}}].

\bibitem{AxionLimits}
C.~O'Hare, ``cajohare/axionlimits: Axionlimits.''
  \url{https://cajohare.github.io/AxionLimits/}, July, 2020.
\newblock 10.5281/zenodo.3932430.

\bibitem{2002clme.book.....G}
H.~{Goldstein}, C.~{Poole} and J.~{Safko}, \emph{{Classical mechanics}} (2002).

\bibitem{Armengaud:2014gea}
E.~Armengaud et~al., \emph{{Conceptual Design of the International Axion
  Observatory (IAXO)}},
  \href{https://doi.org/10.1088/1748-0221/9/05/T05002}{\emph{JINST} {\bfseries
  9} (2014) T05002} [\href{https://arxiv.org/abs/1401.3233}{{\ttfamily
  1401.3233}}].

\bibitem{BREAD:2021tpx}
{\scshape BREAD} collaboration, \emph{{Broadband Solenoidal Haloscope for
  Terahertz Axion Detection}},
  \href{https://doi.org/10.1103/PhysRevLett.128.131801}{\emph{Phys. Rev. Lett.}
  {\bfseries 128} (2022) 131801}
  [\href{https://arxiv.org/abs/2111.12103}{{\ttfamily 2111.12103}}].

\bibitem{Beurthey:2020yuq}
S.~Beurthey et~al., \emph{{MADMAX Status Report}},
  \href{https://arxiv.org/abs/2003.10894}{{\ttfamily 2003.10894}}.

\bibitem{ADMX:2018gho}
{\scshape ADMX} collaboration, \emph{{A Search for Invisible Axion Dark Matter
  with the Axion Dark Matter Experiment}},
  \href{https://doi.org/10.1103/PhysRevLett.120.151301}{\emph{Phys. Rev. Lett.}
  {\bfseries 120} (2018) 151301}
  [\href{https://arxiv.org/abs/1804.05750}{{\ttfamily 1804.05750}}].

\bibitem{ADMX:2018ogs}
{\scshape ADMX} collaboration, \emph{{Piezoelectrically Tuned Multimode Cavity
  Search for Axion Dark Matter}},
  \href{https://doi.org/10.1103/PhysRevLett.121.261302}{\emph{Phys. Rev. Lett.}
  {\bfseries 121} (2018) 261302}
  [\href{https://arxiv.org/abs/1901.00920}{{\ttfamily 1901.00920}}].

\bibitem{ADMX:2019uok}
{\scshape ADMX} collaboration, \emph{{Extended Search for the Invisible Axion
  with the Axion Dark Matter Experiment}},
  \href{https://doi.org/10.1103/PhysRevLett.124.101303}{\emph{Phys. Rev. Lett.}
  {\bfseries 124} (2020) 101303}
  [\href{https://arxiv.org/abs/1910.08638}{{\ttfamily 1910.08638}}].

\bibitem{ADMX:2021mio}
{\scshape ADMX} collaboration, \emph{{Dark matter axion search using a
  Josephson Traveling wave parametric amplifier}},
  \href{https://doi.org/10.1063/5.0122907}{\emph{Rev. Sci. Instrum.} {\bfseries
  94} (2023) 044703} [\href{https://arxiv.org/abs/2110.10262}{{\ttfamily
  2110.10262}}].

\bibitem{ADMX:2021nhd}
{\scshape ADMX} collaboration, \emph{{Search for Invisible Axion Dark Matter in
  the 3.3{\textendash}4.2{\,}{\,}{\ensuremath{\mu}}eV Mass Range}},
  \href{https://doi.org/10.1103/PhysRevLett.127.261803}{\emph{Phys. Rev. Lett.}
  {\bfseries 127} (2021) 261803}
  [\href{https://arxiv.org/abs/2110.06096}{{\ttfamily 2110.06096}}].

\bibitem{Stern:2016bbw}
I.~Stern, \emph{{ADMX Status}},
  \href{https://doi.org/10.22323/1.282.0198}{\emph{PoS} {\bfseries ICHEP2016}
  (2016) 198} [\href{https://arxiv.org/abs/1612.08296}{{\ttfamily
  1612.08296}}].

\bibitem{Crisosto:2019fcj}
N.~Crisosto, P.~Sikivie, N.S.~Sullivan, D.B.~Tanner, J.~Yang and G.~Rybka,
  \emph{{ADMX SLIC: Results from a Superconducting $LC$ Circuit Investigating
  Cold Axions}},
  \href{https://doi.org/10.1103/PhysRevLett.124.241101}{\emph{Phys. Rev. Lett.}
  {\bfseries 124} (2020) 241101}
  [\href{https://arxiv.org/abs/1911.05772}{{\ttfamily 1911.05772}}].

\bibitem{ADMX:2025vom}
{\scshape ADMX} collaboration, \emph{{Search for Axion Dark Matter from 1.1 to
  1.3~GHz with ADMX}}, \href{https://doi.org/10.1103/d7mg-6sqq}{\emph{Phys.
  Rev. Lett.} {\bfseries 135} (2025) 191001}
  [\href{https://arxiv.org/abs/2504.07279}{{\ttfamily 2504.07279}}].

\bibitem{Laverda:2024qjt}
G.~Laverda and J.~Rubio, \emph{{The rise and fall of the Standard-Model Higgs:
  electroweak vacuum stability during kination}},
  \href{https://doi.org/10.1007/JHEP05(2024)339}{\emph{JHEP} {\bfseries 05}
  (2024) 339} [\href{https://arxiv.org/abs/2402.06000}{{\ttfamily
  2402.06000}}].

\bibitem{Laverda:2025pmg}
G.~Laverda and J.~Rubio, \emph{{Higgs-induced gravitational waves: the
  interplay of non-minimal couplings, kination and top quark mass}},
  \href{https://doi.org/10.1007/JHEP08(2025)203}{\emph{JHEP} {\bfseries 08}
  (2025) 203} [\href{https://arxiv.org/abs/2502.04445}{{\ttfamily
  2502.04445}}].

\bibitem{Berera:1995ie}
A.~Berera, \emph{{Warm inflation}},
  \href{https://doi.org/10.1103/PhysRevLett.75.3218}{\emph{Phys. Rev. Lett.}
  {\bfseries 75} (1995) 3218}
  [\href{https://arxiv.org/abs/astro-ph/9509049}{{\ttfamily
  astro-ph/9509049}}].

\bibitem{Mukaida:2012bz}
K.~Mukaida and K.~Nakayama, \emph{{Dissipative Effects on Reheating after
  Inflation}}, \href{https://doi.org/10.1088/1475-7516/2013/03/002}{\emph{JCAP}
  {\bfseries 03} (2013) 002} [\href{https://arxiv.org/abs/1212.4985}{{\ttfamily
  1212.4985}}].

\bibitem{Mukaida:2012qn}
K.~Mukaida and K.~Nakayama, \emph{{Dynamics of oscillating scalar field in
  thermal environment}},
  \href{https://doi.org/10.1088/1475-7516/2013/01/017}{\emph{JCAP} {\bfseries
  01} (2013) 017} [\href{https://arxiv.org/abs/1208.3399}{{\ttfamily
  1208.3399}}].

\bibitem{Mukaida:2013xxa}
K.~Mukaida, K.~Nakayama and M.~Takimoto, \emph{{Fate of $Z_2$ Symmetric Scalar
  Field}}, \href{https://doi.org/10.1007/JHEP12(2013)053}{\emph{JHEP}
  {\bfseries 12} (2013) 053} [\href{https://arxiv.org/abs/1308.4394}{{\ttfamily
  1308.4394}}].

\bibitem{ParticleDataGroup:2020ssz}
{\scshape Particle Data Group} collaboration, \emph{{Review of Particle
  Physics}}, \href{https://doi.org/10.1093/ptep/ptaa104}{\emph{PTEP} {\bfseries
  2020} (2020) 083C01}.

\bibitem{Dunster1990Hankel}
T.M.~Dunster, \emph{Bessel functions of purely imaginary order, with an
  application to second-order linear differential equations having a large
  parameter}, \href{https://doi.org/10.1137/0521055}{\emph{SIAM Journal on
  Mathematical Analysis} {\bfseries 21} (1990) 995}
  [\href{https://arxiv.org/abs/https://doi.org/10.1137/0521055}{{\ttfamily
  https://doi.org/10.1137/0521055}}].

\bibitem{1980Ap&SS..72...33B}
R.~{Broucke}, \emph{{Notes on the central forcer $^{ n }$}},
  \href{https://doi.org/10.1007/BF00642162}{\emph{Astrophysics and Space
  Science} {\bfseries 72} (1980) 33}.

\bibitem{Turner:1983he}
M.S.~Turner, \emph{{Coherent Scalar Field Oscillations in an Expanding
  Universe}}, \href{https://doi.org/10.1103/PhysRevD.28.1243}{\emph{Phys. Rev.
  D} {\bfseries 28} (1983) 1243}.

\end{thebibliography}\endgroup

\end{document}